\documentclass[11pt,a4paper]{article}
\usepackage[T1]{fontenc}
\PassOptionsToPackage{hyperfootnotes=false}{hyperref}
\usepackage{jheppub}
\usepackage{slashed}
\usepackage{tikz-feynman}
\tikzfeynmanset{compat=1.1.0,warn luatex=false}
\usepackage[caption=false]{subfig}
\usepackage{placeins}
\allowdisplaybreaks[1]
\newcommand{\GeV}{\mathrm{GeV}}
\renewcommand{\afterTocRuleSpace}{\clearpage}

\title{\boldmath Invisible dark-sector radiation: analytic resummation for dark photon searches}

\author[a]{Xiangyu Deng,}
\author[a]{Rongfeng Ou,} 
\author[b,a]{Yi Li~\footnote{Corresponding author},}
\author[a]{Mengchao Zhang~\footnote{Corresponding author}}

\affiliation[a]{Department of Physics, College of Physics and Optoelectronic Engineering,
Jinan University, Guangzhou 510632, People’s Republic of China}
\affiliation[b]{School of Physics, Huazhong University of Science and Technology, Wuhan 430074, P. R. China}

\emailAdd{yili2004@hust.edu.cn}
\emailAdd{mczhang@jnu.edu.cn}

\abstract{We derive an analytic resummed description of the missing-mass spectrum in monophoton searches for invisible dark photons at electron--positron colliders. In a dark Abelian Higgs model with a light dark fermion, we construct the spectrum from scalar and fermionic real and virtual contributions without relying on a Monte Carlo shower implementation. 
Sudakov resummation method is used to deal with the collinear divergence. The analytic expressions make the dependence on the dark charges and the logarithmic structure explicit. A simplified reinterpretation of the BaBar monophoton search, accounting for its finite missing-mass resolution, shows that dark-sector radiation can modify the inferred upper limit on the kinetic-mixing parameter by $\mathcal{O}(10\%)$ in the benchmarks considered.
}

\begin{document}
\maketitle

\section{Introduction \label{sec:intro}}

Collider searches for dark matter (DM) and other invisible dark sector particles use missing-momentum and recoil observables~\cite{Abercrombie:2015wmb}. 
At hadron colliders, these include missing transverse momentum $E_T^{\rm miss}$ and the stransverse mass $M_{T2}$ for pair-decay topologies~\cite{Lester:1999MT2}. 
Electron--positron colliders also use missing-mass reconstruction~\cite{Birkedal:2004Recoil}.
For a monophoton final state at centre-of-mass energy \(\sqrt s\), the squared missing mass \(M_X^2\equiv s-2\sqrt sE_\gamma\) gives the invariant mass squared of the invisible system.
Beyond discovery, $M_{T2}$ endpoints and kinks can constrain parent and invisible-daughter masses~\cite{Cho:2007GluinoMT2,Cho:2007Kink,Burns:2008SubsystemMT2,Konar:2009AsymmetricMT2}, while missing-momentum and recoil-mass spectra probe DM and mediator mass scales under specified production assumptions~\cite{Franzosi:2015Mass,Andersen:2013Spectroscopy}. 
Reliable differential predictions are therefore central to extracting particle properties from an invisible signal.

A particularly well-studied dark sector particle is the dark photon (DP), $A'$, the gauge boson of a dark $U(1)'$ symmetry~\cite{Okun:1982Paraphotons,Galison:1983Mixing,Holdom:1986Mixing,Fabbrichesi:2020Review}. Kinetic mixing, parametrized by $\varepsilon$, couples the massive DP to the Standard Model (SM) electromagnetic current, and thus enables the DP to be probed in a wide variety of experiments~\cite{BaBar:2014zli,BaBar:2017tiz,BESIII:2017fwv,Merkel:2014avp,NA64:2016oww,NA64:2017vtt,Banerjee:2019pds,LHCb:2017trq,LHCb:2019vmc,CMS:2019buh,FASER:2023tle,KLOE-2:2011hhj,KLOE-2:2012lii,KLOE-2:2014qxg,KLOE-2:2016ydq,Zhang:2019wnz,Belle-II:2018jsg,NA482:2015wmo,Andreas:2013lya,Gninenko:2013rka,SHiP:2021nfo,DarkSHINE:2022mak,Duarte:2022feb,Ilten:2015hya,Ilten:2016tkc,Feng:2017uoz}.
These experiments already exclude broad regions of the $(m_{A'},\varepsilon)$ plane under specified decay assumptions.
The two-parameter plane $(m_{A'},\varepsilon)$ has consequently become the standard presentation of DP constraints.

\begin{figure}[!htbp]
\centering
\includegraphics[width=1.0\textwidth]{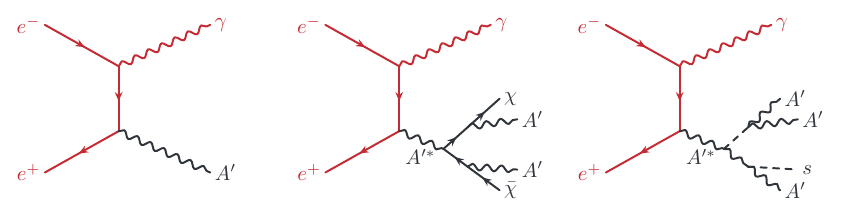}
\caption{Left: The process considered in most invisible DP searches, with $M^2_X = m^2_{A'}$ at Born level. Middle \& Right: Dark sector radiation channels considered in this work, which generate a continuum in \(M_X^2\).}
\label{fig:FD}
\end{figure}

That two-parameter description does not, however, fix the dynamics of the dark sector. 
In an invisible-decay realization, light dark charged states (labeled by $\chi$) provide the dominant decay products of the DP; when the DP mass arises from spontaneous symmetry breaking, a dark Higgs (labeled by $s$) is also present~\cite{Pospelov:2007mp,Batell:2009yf,Kim:2016fdv}~\footnote{The presence of a dark Higgs motivates dark Higgsstrahlung searches at BaBar, Belle, KLOE, and Belle II~\cite{BaBar:2012bkw,Jaegle:2015fme,KLOE-2:2015nli,Belle-II:2022jyy}, including proposals with displaced visible decays or other signals~\cite{Acanfora:2025ikj,Falkowski:2010cm,Biswas:2022vkq,ATLAS:2022hzh,ATLAS:2023zxo,ATLAS:2024darkgamma,ATLAS:2025leptonjets,Yan:2022npz}. These channels test different decay patterns from the fully invisible final states we study in this work. }. 
Their masses and couplings are not determined by $m_{A'}$ and $\varepsilon$. 
Associated production and off-shell channels can populate an invisible continuum alongside the on-shell DP contribution~\cite{Cheung:2024oxh,Li:2025tlg}, and logarithmically enhanced dark final-state radiation (FSR) can reshape the spectrum~\cite{Kim:2016fdv,Chen:2018uii} (see Fig.~\ref{fig:FD} for illustration). 
Thus, even when every dark particle escapes detection, the monophoton recoil can retain information about the dark sector dynamics behind the portal.

In this work we derive an analytic, resummed prediction for $e^+e^- \to \gamma + X$ in a dark Abelian Higgs model with a light dark fermion. 
Multiple dark emissions are commonly treated as dark showers using Monte Carlo (MC) event generators~\cite{Carloni:2010tw,Carloni:2011kk,Buschmann:2015awa,Bierlich:2022pfr,Kim:2016fdv,Chen:2018uii,Li:2021MultiScalar,Schwaller:2015EmergingJets,Cohen:2015SemiVisibleJets,Cohen:2017DarkShowers,Park:2017DarkJets,Cohen:2020JetSubstructure,Lu:2023DarkQCD,Chen:2025GluonJets,Zhang:2021LeptophilicADM,Lu:2025Leptophilic,Chigusa:2022act,Li:2022BoostedSplitting,Li:2023SecludedShowers}.
Our calculation instead connects the underlying amplitudes to the missing-mass spectrum through resummation and matching to the fixed-order continuum, with inclusive next-to-leading-order (NLO) normalization. 
The analytic results show explicitly how the dark charges and couplings shape the missing-mass spectrum, providing insight into dark-sector dynamics that complements MC studies.
In particular, the primary splitting kernels of the neutral DP have no soft poles. Within our approximation, the resummed endpoint spectrum is governed by single collinear logarithms, rather than the familiar soft--collinear double logarithms in QCD and QED showers. 
Finally, we perform a simplified reinterpretation of the BaBar monophoton search to examine how these radiative corrections affect the inferred limits on the kinetic-mixing parameter.

The paper is organized as follows. 
Sec.~\ref{sec:model} defines the model we want to study. 
Secs.~\ref{sec:Xsection} and~\ref{sec:wo_res} present the inclusive NLO calculation and the fixed-order spectrum. 
Sec.~\ref{sec:resum} is devoted to the resummation. 
Sec.~\ref{sec:real_collider} combines matching, detector response, and the BaBar limit reinterpretation. 
We conclude this work in Sec.~\ref{sec:conclu}.

\section{Dark Abelian Higgs model with invisible final states \label{sec:model}}

We use a dark Abelian Higgs model as a renormalizable benchmark dark sector model in which the origin of the DP mass and its invisible decay channel are both explicit. 
Unlike a Stueckelberg realization without a light physical scalar, the Higgs realization relates the scalar and vector masses to the symmetry-breaking scale and the corresponding couplings. 
The general role of perturbative unitarity in broken gauge theories~\cite{Lee:1977yc,Lee:1977eg}, and explicit bounds on the dark-Higgs mass in a $U(1)'$ self-interacting DM model~\cite{Li:2024wqj}, motivate retaining the scalar rather than assuming an unrestricted decoupling limit. 
The separation between a weak production portal and stronger dark sector interactions is also realized in ultraviolet constructions~\cite{Rizzo:2024bhn} and secluded models~\cite{Pospelov:2007mp,Arkani-Hamed:2008hhe,Pospelov:2008jd}.
A light dark fermion $\chi$ is introduced so that the DP decays predominantly invisibly; the dark Higgs also decays into invisible dark-sector states.
This work focuses on collider phenomenology. 
For cosmological and direct-detection studies of this model, see Refs.~\cite{Alves:2009nf,Hisano:2003ec,March-Russell:2008lng,Chen:2023rrl,Baldes:2017gzw,Graesser:2011wi,Ibe:2019ena,Arkani-Hamed:2008kxc,Cirelli:2008pk,Cholis:2008wq,Cholis:2008qq,PandaX-II:2016vec,LUX:2016ggv}.

The Lagrangian relevant to the present calculation is
\begin{equation} \label{eq:2_01} 
\mathcal{L}_{\text{dark}} = ( D_\mu S )^\dagger D^\mu S - \frac{1}{4} F'_{\mu\nu} F'^{\mu\nu} + \mu_S^2 S^\dagger S - \frac{1}{4} \lambda (S^\dagger S)^2 + \bar{\chi} ( i \slashed{D} - m_\chi ) \chi - \varepsilon e J^{\text{EM}}_\mu A'^{\mu}  
\end{equation}
with $S$ the complex dark Higgs and $F'_{\mu\nu} = \partial_\mu A'_\nu - \partial_\nu A'_\mu$ the DP field strength.
$D_\mu = \partial_{\mu} - ig' Q_i A'_\mu$ is the dark $U(1)'$ covariant derivative, and $g'$ is the dark gauge coupling.
The $U(1)'$ charges of $S$ and $\chi$ are $Q_S$ and $Q_\chi$, respectively.  
$J^{\text{EM}}_\mu$ stands for the SM electromagnetic current, and $\varepsilon$ is the kinetic mixing parameter.
$e$ is the SM electric charge. 
We further define the dark fine structure constant $\alpha' \equiv g'^2/4\pi$ for later convenience. 
The negative mass square $-\mu_S^2$, the self-coupling $\lambda$, and the dark fermion mass $m_\chi$ are other parameters. 
The coupling for the possible Higgs portal $(H^\dagger H)(S^\dagger S)$ is assumed to be negligibly small in this work.

After spontaneous symmetry breaking (SSB), $S$ can be expanded as
\begin{equation} \label{eq:2_02} 
S = \frac{1}{\sqrt{2}} \left( v + s + ia \right) ,
\end{equation}
where $v \equiv 2\mu_S/\sqrt{\lambda}$ is the vacuum expectation value (VEV) of $S$, and $s$ the physical dark Higgs. 
The squared masses for the DP and the dark Higgs are
\begin{equation} \label{eq:2_03} 
m^2_{A'} = Q_{S}^2 g'^2 v^2  \ , \ m^2_s = \frac{1}{2} \lambda v^2
\end{equation}
We retain $Q_S$ and $Q_\chi$ explicitly in order to separate the scalar and fermionic contributions to the radiation spectrum and the inclusive rate. 
This parameterization explores their relative weights within the model. 
Thus the input parameters in this work can be chosen as
\begin{equation} \label{eq:2_04} 
\varepsilon \ , \ m_\chi  \ , \ m_{A'} \ , \ m_s \ , \ \alpha' \  , \ Q_S \ , \ Q_\chi
\end{equation}
We take \(m_\chi\ll m_{A'},m_s\) and assume that dark decays dominate, so that the relevant final states are invisible~\footnote{In this dark sector model $s$ decays to the $\chi\bar{\chi}\chi\bar{\chi}$ final state via $A'$.}.
In our following analysis we also take the limit of $m_\chi\to 0$.
And $m_s$ might be traded for the squared mass ratio $r\equiv {m_{A'}^2}/{m_s^2}$ in the following analysis. 

\section{NLO inclusive cross section \label{sec:Xsection}}

In this section we calculate the NLO inclusive cross section for the process $e^+e^-\to \gamma + X$, where $X$ in the final state can be $A'$ (on-shell DP), $A'+s$, or $\bar{\chi}\chi$ (off-shell DP). 
This inclusive process corresponds to the ``mono-photon + $\slashed{E}$'' signal at lepton colliders. 
The dark-Higgs real and virtual corrections to the inclusive cross section have been given in Ref.~\cite{Zheng:2026mji}, while the contribution from the dark fermion is new. 
Throughout this paper, NLO refers to corrections in the dark coupling $\alpha'$, not to a complete NLO treatment of SM electromagnetic radiation. 
A subtle point is that the three-body channel $e^+e^-\to \gamma\chi\bar\chi$ contains an intermediate $A'$ resonance region. 
Since the production of an on-shell $A'$ is already counted by the Born process $e^+e^-\to\gamma A'$, the resonant part of the three-body rate should not be counted again as a genuine real-emission correction. 
Therefore, throughout this section we define $\sigma_R^{(\chi)}$ as a \emph{resonance-subtracted} three-body cross section. 
With this convention, we define $\sigma_{\rm NLO}$ as the total cross section including both the leading-order (LO) and NLO contributions:
\begin{equation} \label{eq:3_01}
\sigma_{\rm NLO} =
\sigma_0 + \sigma_V^{(s)}+\sigma_V^{(\chi)}+\sigma_R^{(s)}+\sigma_R^{(\chi)}.
\end{equation}
Here $\sigma_0$ denotes the LO two-body contribution $e^+e^-\to\gamma A'$, while the genuine NLO correction is given by $\sigma_{\rm NLO}-\sigma_0$. 
Throughout this work, $\sigma_0$ and $\sigma_{\rm LO}$ are used interchangeably for this LO contribution, whereas $\sigma_V$ and $\sigma_R$ denote the virtual and real-emission corrections, respectively. 
We will show that the logarithmic divergences appearing in the separate real and virtual pieces cancel in the inclusive combination.

Before the concrete calculation, here we list the Feynman rules that are relevant for the present NLO analysis (in addition to the standard QED vertex):
\begin{equation}\label{eq:3_02}
\begin{gathered}
\begin{tikzpicture}[scale=0.7, transform shape ,baseline=(a)]
    \begin{feynman}
    \vertex(a){\(A'^\mu\)};
    \vertex[right=of a](b);
    \vertex[above right=of b](c){\(s\)};
    \vertex[below right=of b](d){\(A'^\nu\)};
    \diagram*{
    (a)--[boson](b)--[boson](d),
    (b)--[scalar](c),
    };
    \end{feynman}
    \end{tikzpicture}
    =2i  Q_S g' m_{A'} g_{\mu\nu} \ \ ,  \ \ 
    \begin{tikzpicture}[scale=0.7, transform shape ,baseline=(a)]
    \begin{feynman}
    \vertex(a){\(A'^\mu\)};
    \vertex[right=of a](b);
    \vertex[above right=of b](c){\(e^-\)};
    \vertex[below right=of b](d){\(e^+\)};
    \diagram*{
    (a)--[boson](b),
    (b)--[fermion](c),
    (d)--[fermion](b),
    };
    \end{feynman}
    \end{tikzpicture}
    = -i \varepsilon e \gamma^\mu \ \ ,  \ \ \\[3pt]
    \begin{tikzpicture}[scale=0.7, transform shape ,baseline=(a)]
    \begin{feynman}
    \vertex(a){\(A'^\mu\)};
    \vertex[right=of a](b);
    \vertex[above right=of b](c){\(\chi\)};
    \vertex[below right=of b](d){\(\bar{\chi}\)};
    \diagram*{
    (a)--[boson](b),
    (b)--[fermion](c),
    (d)--[fermion](b),
    };
    \end{feynman}
    \end{tikzpicture}
    = i Q_\chi g' \gamma^\mu
\end{gathered}
\end{equation}
Thus the coupling power counting is
\begin{equation} \label{eq:3_03}
\sigma_0 \propto \varepsilon^2 \alpha^2 \, , \qquad
\sigma_V^{(s)},\sigma_V^{(\chi)},\sigma_R^{(s)},\sigma_R^{(\chi)} \propto \varepsilon^2 \alpha' \alpha^2
\end{equation}
with $\alpha\equiv e^2/4\pi$ the SM fine-structure constant.

\subsection{Amplitudes \label{sec:Xsection_1}}

In this subsection we present the amplitudes entering the inclusive cross section.

For the LO process $ e^+(p_2) + e^-(p_1) \to \gamma(k) + A'(q)$, the amplitude is
\begin{equation} \label{eq:3_04}
 i\mathcal{M}_0 = - i\varepsilon e^2  \bar{v}(p_2) \left[ \slashed{\epsilon}(q) \frac{1}{\slashed{p}_1-\slashed{k}-m_e} \slashed{\epsilon}(k) +
 \slashed{\epsilon}(k)  \frac{1}{\slashed{k}-\slashed{p}_2-m_e} \slashed{\epsilon}(q) \right] u(p_1).
\end{equation}
Here $\epsilon(k)$ and $\epsilon(q)$ denote the outgoing polarization vectors of the photon and DP, respectively (complex conjugates have been implied).
Spin indices of the fermion wave functions are suppressed for conciseness.  

For the real-emission process $e^+(p_2) + e^-(p_1) \to \gamma(k) + A'(Q)^\ast \to \gamma(k) + A'(k_1) + s(k_2)$, the amplitude is
\begin{equation}\label{eq:3_05}
\begin{aligned}
 i\mathcal{M}^{(s)}_R ={}& -i\varepsilon e^2\frac{2Q_S g' m_{A'}}{Q^2-m_{A'}^2}\bar v(p_2)\\
 &\times\left[\slashed{\epsilon}(k_1)\frac{1}{\slashed{p}_1-\slashed{k}-m_e}\slashed{\epsilon}(k)
 +\slashed{\epsilon}(k)\frac{1}{\slashed{k}-\slashed{p}_2-m_e}\slashed{\epsilon}(k_1)\right]u(p_1),
\end{aligned}
\end{equation}
with $Q\equiv k_1 + k_2$ the momentum carried by the intermediate $A'$.

For the real-emission process $e^+(p_2) + e^-(p_1) \to \gamma(k) + A'(Q)^\ast \to \gamma(k) + \chi(k_1) + \bar{\chi}(k_2)$, the amplitude is
\begin{equation}\label{eq:3_06}
\begin{aligned}
 i\mathcal{M}^{(\chi)}_R ={}& -i\varepsilon e^2 \frac{Q_\chi g'}{Q^2-m_{A'}^2}\bar{v}(p_2) \\
 &\times\left[\gamma^\mu\frac{1}{\slashed{p}_1-\slashed{k}-m_e}\slashed{\epsilon}(k)
 +\slashed{\epsilon}(k)\frac{1}{\slashed{k}-\slashed{p}_2-m_e}\gamma^\mu\right]
 u(p_1)\bar{u}(k_1)\gamma_\mu v(k_2).
\end{aligned}
\end{equation}

The virtual correction to the two-body amplitude is induced by the DP self-energy. 
Since the one-loop self-energy receives both the dark-Higgs contribution and the dark-fermion contribution, it is convenient to write
\begin{equation} \label{eq:3_07}
 i\mathcal{M}_V =  i\mathcal{M}^{(s)}_V +i\mathcal{M}^{(\chi)}_V  = \mathcal{C}^{(s)}_V  i\mathcal{M}_0 + \mathcal{C}^{(\chi)}_V  i\mathcal{M}_0.
\end{equation}
The relevant self-energy is the transverse one. 
Indeed, after summing the two electron-line diagrams, the current coupled to the DP is conserved, so that the longitudinal term proportional to $p^\mu p^\nu$ does not contribute. 
The transverse self-energy of the DP can be written as
\begin{equation} \label{eq:3_08}
 i\Pi_{A'}^{\mu\nu}(p) = i\left(-g^{\mu\nu}+\frac{p^\mu p^\nu}{p^2}\right)\Pi_{A'}(p^2),
\end{equation}
and thus the virtual factors are obtained from the wave-function renormalization as
\begin{equation} \label{eq:3_09}
\mathcal{C}^{(s)}_V =  \frac{1}{2} \frac{d}{d p^2} \Pi^{(s)}_{A'}(p^2) \Big|_{p^2=m_{A'}^2}   \ , \ 
\mathcal{C}^{(\chi)}_V =  \frac{1}{2} \frac{d}{d p^2} \Pi^{(\chi)}_{A'}(p^2) \Big|_{p^2=m_{A'}^2}.
\end{equation}

By using the DP self-energy given in a previous study~\cite{Dudal:2019aew}, the dark-Higgs contribution can be written in the closed form~\cite{Zheng:2026mji}
\begin{equation}\label{eq:3_10}
\begin{aligned}
 \mathcal{C}_V^{(s)}={}&\frac{Q_S^2\alpha'}{4\pi}\frac{1}{36\sqrt{4r-1}}\Bigg[
 \sqrt{4r-1}\Bigg(6\ln\frac{m_{A'}^2}{\mu^2}+62-\frac{36}{r}+\frac{12}{r^2}\\
 &\qquad+\left(-36+\frac{54}{r}-\frac{27}{r^2}+\frac{6}{r^3}\right)\ln r\Bigg)\\
 &+6\left(36-\frac{32}{r}+\frac{13}{r^2}-\frac{2}{r^3}\right)
 \left(-\tan^{-1}\frac{1}{\sqrt{4r-1}}+\tan^{-1}\frac{1-2r}{\sqrt{4r-1}}\right)\Bigg].
\end{aligned}
\end{equation}
where $r\equiv m_{A'}^2/m_s^2$ is the ratio of squared masses, and $\mu$ the renormalization scale. 
The analytic expression shown here corresponds to the branch $r>1/4$ used in the numerical benchmarks. 
For other regions the result should be understood through the corresponding analytic continuation of the loop functions.

After dimensional regularization and ultraviolet subtraction, using the scale convention of Ref.~\cite{Dudal:2019aew},
the finite dark-fermion contribution is
\begin{equation} \label{eq:3_11}
\Pi^{(\chi)}_{A'}(p^2) = \frac{2\alpha'Q_\chi^2}{\pi}\,p^2\int_0^1 dx\,x(1-x) \ln\frac{m_\chi^2-x(1-x)p^2-i0^+}{\mu^2}\, .
\end{equation}
with $\mu$ the renormalization scale. 
Taking one derivative with respect to $p^2$ gives
\begin{align} \label{eq:3_12}
\mathcal{C}_V^{(\chi)}
&= \frac{\alpha'Q_\chi^2}{\pi}\int_0^1 dx\left[
 x(1-x)\ln\frac{m_\chi^2-x(1-x)m_{A'}^2-i0^+}{\mu^2}
 -\frac{m_{A'}^2x^2(1-x)^2}{m_\chi^2-x(1-x)m_{A'}^2-i0^+}
\right].
\end{align}
For the kinematic region relevant to invisible decays, $m_{A'}>2m_\chi$, the above Feynman-parameter integral can be carried out analytically:
\begin{align} \label{eq:3_13}
 \mathcal{C}_V^{(\chi)} &= \frac{\alpha' Q_\chi^2}{\pi}
\left[ \frac{1}{6}\ln\frac{m_\chi^2}{\mu^2}
-\frac{3\beta_\chi^2+1}{36}
+\frac{\beta_\chi^4+3}{24\beta_\chi} L_\chi \right]\, ,
\end{align}
with
\begin{equation} \label{eq:3_14}
 \beta_\chi \equiv \sqrt{1-\frac{4m_\chi^2}{m_{A'}^2}} \ , \
L_\chi \equiv \ln\frac{1+\beta_\chi}{1-\beta_\chi} - i\pi.
\end{equation}
In particular, in the practically relevant limit $m_\chi\to 0$, one has
\begin{equation} \label{eq:3_15}
 \mathcal{C}_V^{(\chi)} \xrightarrow[m_\chi\to 0]{}
 \frac{\alpha'Q_\chi^2}{\pi}
 \left[
 \frac{1}{6}\ln\frac{m_{A'}^2}{\mu^2}-\frac{1}{9}-\frac{i\pi}{6}
 \right].
\end{equation}

\subsection{Cross sections and resonance subtraction \label{sec:Xsection_2}}

We now convert the above amplitudes into cross sections. 
The decomposition introduced in Eq.~(\ref{eq:3_01}) is realized at the phase-space level as
\begin{equation}  \label{eq:3_16}
 \sigma_{\rm NLO}=\sigma_0 +\sigma_V^{(s)} +\sigma_V^{(\chi)} +\sigma_R^{(s)} +\sigma_R^{(\chi)}\, ,
\end{equation}
with
\begin{align}   \label{eq:3_17}
 \sigma_{\rm NLO} &= \frac{1}{4\sqrt{(p_1\cdot p_2)^2-m_e^4}}
\Bigg[ \int d\Pi_2 \left(  \overline{|\mathcal{M}_0|^2} +
 \overline{\mathcal{M}_0\mathcal{M}_V^{\dagger}}  + \overline{\mathcal{M}_0^{\dagger}\mathcal{M}_V} \right) \nonumber \\
& + \int d\Pi_3^{(s)}  \overline{|\mathcal{M}^{(s)}_R|^2} + \int d\Pi_3^{(\chi)}  \overline{|\mathcal{M}^{(\chi)}_R|^2} \Bigg].
\end{align}
Here $d\Pi_2$, $d\Pi_3^{(s)}$, and $d\Pi_3^{(\chi)}$ denote the corresponding two-body and three-body phase space measures, but $d\Pi_3^{(\chi)}$ should not contain the resonance region as we will explain soon. 
The spin-averaged squared amplitude is defined by
\begin{equation}  \label{eq:3_18}
 \overline{|\mathcal{M}|^2} = \frac{1}{4}\sum_{\rm spin}|\mathcal{M}|^2\, .
\end{equation}
Since $ i\mathcal{M}_V = (\mathcal{C}^{(s)}_V + \mathcal{C}^{(\chi)}_V  ) i\mathcal{M}_0$, the virtual corrections factorize as
\begin{align}  \label{eq:3_19}
\sigma_V^{(s)} +\sigma_V^{(\chi)} &= \left( \mathcal{C}^{(s)}_V + \mathcal{C}^{(\chi)}_V
+ \mathcal{C}^{(s)\ast}_V + \mathcal{C}^{(\chi)\ast}_V   \right) \sigma_0 \nonumber \\
&=  \left( 2{\rm Re }\, \mathcal{C}^{(s)}_V + 2{\rm Re }\, \mathcal{C}^{(\chi)}_V \right) \sigma_0 \, . 
\end{align}

The leading-order two-body cross section for $e^+e^-\to \gamma A'$ is 
\begin{equation}\label{eq:3_20}
\begin{aligned}
 \sigma_0={}&\frac{4\pi\varepsilon^2\alpha^2}{\beta_e^2s^2(s-m_{A'}^2)}
 \Big[2\left(m_{A'}^4-4m_{A'}^2m_e^2-8m_e^4+4m_e^2s+s^2\right)
 \operatorname{arctanh}\beta_e\\
 &\hspace{6em}-\beta_e\left(m_{A'}^4+s(s+4m_e^2)\right)\Big]\, .
\end{aligned}
\end{equation}
with $\beta_e\equiv \sqrt{1-\frac{4m_e^2}{s}}$.
Here we need to keep the electron mass to avoid the QED collinear divergence. 

For later use it is convenient to define the off-shell two-body cross section, namely $\sigma_0$ with $m^2_{A'}$ replaced by $Q^2$:
\begin{align} \label{eq:3_21}
& \hat\sigma_0(s,Q^2)
= \\ \nonumber
& \frac{4\pi\varepsilon^2\alpha^2}{\beta_e^2s^2(s-Q^2)}
\left[
2\left(Q^4-4Q^2m_e^2-8m_e^4+4m_e^2s+s^2\right)
\operatorname{arctanh}\beta_e
-\beta_e\left(Q^4+s(s+4m_e^2)\right)
\right].
\end{align}
By construction, $\sigma_0 = \hat\sigma_0(s,m_{A'}^2)$.

The dark-Higgs real-emission cross section can be reduced to a one-dimensional integral over the invariant mass squared $Q^2=(k_1+k_2)^2$ of the intermediate $A'$:
\begin{equation}\label{eq:3_22}
\begin{aligned}
 \sigma_R^{(s)}={}&\int_{(m_{A'}+m_s)^2}^{q_{\rm cut}^2}dQ^2\,
 \hat\sigma_0(s,Q^2)\frac{Q_S^2\alpha' m_{A'}^2}{\pi Q^2(Q^2-m_{A'}^2)^2}\\
 &\times\lambda^{1/2}(Q^2,m_{A'}^2,m_s^2)
 \left[1+\frac{\lambda(Q^2,m_{A'}^2,m_s^2)}{12m_{A'}^2Q^2}\right]\, .
\end{aligned}
\end{equation}
with the $\lambda$ function defined as
\begin{equation}  \label{eq:3_23}
 \lambda(x,y,z)=x^2+y^2+z^2-2xy-2xz-2yz,
\end{equation}
and
\begin{equation}  \label{eq:3_24}
 q_{\rm cut}^2 \equiv s-2 E_\gamma^{\rm cut} \sqrt{s}.
\end{equation}
Here \(E_\gamma^{\rm cut}\) is the minimal value for the final state photon energy, which is introduced to avoid the soft QED divergence and is also required for signal tagging at a collider. 

For the dark-fermion channel we first introduce the width-regulated three-body rate
\begin{equation} \label{eq:3_25}
 \bar\sigma_R^{(\chi)} =
 \int_{4m_\chi^2}^{q_{\rm cut}^2} dQ^2\, \hat\sigma_0(s,Q^2)
 \frac{1}{\pi}  \frac{\sqrt{Q^2}\,\Gamma_{\chi}(Q^2)}{(Q^2-m_{A'}^2)^2+m_{A'}^2\Gamma_{A'}^2},
\end{equation}
with the off-shell partial width
\begin{equation} \label{eq:3_26}
 \Gamma_{\chi}(Q^2)  =
 \frac{\alpha'Q_\chi^2}{3}\, \sqrt{Q^2}
 \left(1+\frac{2m_\chi^2}{Q^2}\right)  \sqrt{1-\frac{4m_\chi^2}{Q^2}}\, .
\end{equation}
Here $\Gamma_{A'}$ is the on-shell total width of $A'$. 
In the invisible-decay setup considered in this work, it is natural to take $\Gamma_{A'}\simeq\Gamma_\chi(m_{A'}^2)$~\footnote{The DP can also decay to light SM fermion pairs via kinetic mixing, but the corresponding width is much smaller than $\Gamma_\chi$.}.

The reason for introducing a width-regulated form is that the region $Q^2\simeq m_{A'}^2$ corresponds to the production of a nearly on-shell $A'$ followed by its invisible decay. 
This part is already represented by the two-body process $e^+e^-\to\gamma A'$, and therefore it should not be counted again as a genuine real-emission correction. 
To subtract this contribution exactly over the physical integration range, we define the normalized resonant profile
\begin{equation} \label{eq:3_27}
 \mathcal{P}_{\rm res}(Q^2) =
 \frac{1}{\mathcal{N}_{\rm res}}
 \frac{1}{\pi}
 \frac{m_{A'}\Gamma_{A'}}{(Q^2-m_{A'}^2)^2+m_{A'}^2\Gamma_{A'}^2}\, ,
\end{equation}
with the normalization factor given by
\begin{align} \label{eq:3_28}
 \mathcal{N}_{\rm res}
 &=
 \int_{4m_\chi^2}^{q_{\rm cut}^2} dQ^2\,
 \frac{1}{\pi}
 \frac{m_{A'}\Gamma_{A'}}{(Q^2-m_{A'}^2)^2+m_{A'}^2\Gamma_{A'}^2}  \nonumber \\
 &=
 \frac{1}{\pi}
 \left[ \tan^{-1}\frac{q_{\rm cut}^2-m_{A'}^2}{m_{A'}\Gamma_{A'}} -  \tan^{-1}\frac{4m_\chi^2-m_{A'}^2}{m_{A'}\Gamma_{A'}} \right]\, .
\end{align}
Then the subtraction term is chosen as
\begin{equation} \label{eq:3_29}
 \sigma_0\, \mathrm{Br}(A'\to\chi\bar\chi)\, \mathcal{P}_{\rm res}(Q^2)\, ,
\end{equation}
which satisfies
\begin{equation} \label{eq:3_30}
 \int_{4m_\chi^2}^{q_{\rm cut}^2} dQ^2\, \sigma_0\, \mathrm{Br}(A'\to\chi\bar\chi)\, \mathcal{P}_{\rm res}(Q^2)
 = \sigma_0\,\mathrm{Br}(A'\to\chi\bar\chi)
\end{equation}
exactly. 
Therefore, in the remainder of this paper we define the real correction from the dark-fermion channel as the resonance-subtracted rate
\begin{equation} \label{eq:3_31}
 \sigma_R^{(\chi)} =
 \int_{4m_\chi^2}^{q_{\rm cut}^2} dQ^2
 \left[ \hat\sigma_0(s,Q^2) \frac{1}{\pi}
 \frac{\sqrt{Q^2}\,\Gamma_{\chi}(Q^2)}{(Q^2-m_{A'}^2)^2+m_{A'}^2\Gamma_{A'}^2}
 - \sigma_0\, \mathrm{Br}(A'\to\chi\bar\chi)\, \mathcal{P}_{\rm res}(Q^2) \right]\, .
\end{equation}
When $\mathrm{Br}(A'\to\chi\bar\chi)\simeq 1$, this normalized pole-profile subtraction removes one copy of the LO pole contribution over the chosen $Q^2$ window, so that $\sigma_R^{(\chi)}$ behaves as a genuine real-emission correction.  

\subsection{IR safety analysis \label{sec:Xsection_3}}

In this subsection we analyze the IR structure of the inclusive cross section. The cancellation of mass singularities between degenerate real and virtual contributions is governed by the Kinoshita--Lee--Nauenberg theorem~\cite{Kinoshita:1962ur,Lee:1964is}; an inclusive cancellation does not by itself determine the spectrum in a resolved invariant-mass variable.
The $s$ and $\chi$ channels can be treated in parallel, but the corresponding kernels are different. 
For the dark-Higgs channel, the logarithmic singularity is the same as the one studied in our previous fixed-order work~\cite{Zheng:2026mji}. 
For the dark-fermion channel, since we are primarily interested in the invisible-decay setup with $m_\chi\ll m_{A'}$, it is convenient to directly work in the approximation $m_\chi\to 0$ from this point on. 
The finite-$m_\chi$ result can be straightforwardly recovered, and it does not modify the qualitative conclusions below as long as \(m_\chi\ll m_{A'}\).

\paragraph{Dark-Higgs channel.}
Although this part has already been discussed in our previous work~\cite{Zheng:2026mji}, for completeness we briefly summarize the extraction here. 
Keeping only the logarithmic term of $\mathcal C_V^{(s)}$ in the limit $m_{A'}\to 0$ with fixed $r\equiv m_{A'}^2/m_s^2$, one finds the IR-divergent part of $ \sigma^{(s)}_V$:
\begin{equation} \label{eq:3_32}
 \sigma^{(s)}_{V,{\rm IR}} =
 2\,\mathrm{Re}\,\mathcal C^{(s)}_{V,{\rm IR}}\,\sigma_0=
 \frac{Q_S^2\alpha'}{12\pi}
 \ln\frac{m_{A'}^2}{\mu^2}\,\sigma_0\, .
\end{equation}
To extract the real IR singularity, we start from Eq.~(\ref{eq:3_22}) and define the dimensionless variable
\begin{equation} \label{eq:3_33}
 x \equiv \frac{Q^2}{m_{A'}^2}
\end{equation}
in the limit $m_{A'}\to 0$ with fixed $r=m_{A'}^2/m_s^2$. 
Then Eq.~(\ref{eq:3_22}) can be rewritten as
\begin{equation} \label{eq:3_34}
 \sigma_R^{(s)} =
 \frac{Q_S^2\alpha'}{\pi}
 \int_{x_{\rm min}}^{x_{\rm max}} dx\,
 \hat\sigma_0(s,m_{A'}^2x)
 \frac{\sqrt{\Lambda_r(x)}}{x(x-1)^2}
 \left[ 1+\frac{\Lambda_r(x)}{12x} \right]\, ,
\end{equation}
with
\begin{equation} \label{eq:3_35}
 x_{\rm min}=\left(1+\frac{1}{\sqrt r}\right)^2 \ , \
 x_{\rm max}=\frac{q_{\rm cut}^2}{m_{A'}^2}\, ,
\end{equation}
and
\begin{equation} \label{eq:3_36}
 \Lambda_r(x)=x^2+1+\frac{1}{r^2}-2x-\frac{2x}{r}-\frac{2}{r}\, .
\end{equation}
The logarithmic contribution comes from the large-$x$ region. 
Expanding the kernel in powers of $1/x$, one has
\begin{align} \label{eq:3_37}
 \sqrt{\Lambda_r(x)}
 &=
 x-\left(1+\frac{1}{r}\right)+{\cal O}\!\left(\frac{1}{x}\right)\, ,
 \nonumber \\
 1+\frac{\Lambda_r(x)}{12x}
 &=
 1+\frac{x}{12}-\frac{1}{6}\left(1+\frac{1}{r}\right)+{\cal O}\!\left(\frac{1}{x}\right)\, ,
\end{align}
and therefore
\begin{equation} \label{eq:3_38}
 \frac{\sqrt{\Lambda_r(x)}}{x(x-1)^2}
 \left[  1+\frac{\Lambda_r(x)}{12x} \right]
 =  \frac{1}{12x}+{\cal O}\!\left(\frac{1}{x^2}\right).
\end{equation}
Meanwhile, in the same limit one may expand the off-shell two-body cross section as
\begin{equation} \label{eq:3_39}
\hat\sigma_0(s,m_{A'}^2x)=\sigma_0\left[1+{\cal O}\!\left(\frac{m_{A'}^2(x-1)}{s}\right)\right].
\end{equation}
This expansion is used in the overlap region $m_{A'}^2x\ll s$.  The second term only gives a finite contribution to the logarithmic extraction, so the IR-divergent part for the real emission process is
\begin{equation} \label{eq:3_40}
 \sigma^{(s)}_{R,{\rm IR}}
 =
 \frac{Q_S^2\alpha'}{12\pi}\,\sigma_0
 \int^{x_{\rm max}} dx\,\frac{1}{x}
 =
 -\frac{Q_S^2\alpha'}{12\pi}
 \ln\frac{m_{A'}^2}{q_{\rm cut}^2}\,\sigma_0\, .
\end{equation}
Combining Eqs.~(\ref{eq:3_32}) and (\ref{eq:3_40}), we obtain
\begin{equation} \label{eq:3_41}
 \sigma^{(s)}_{V,{\rm IR}}+\sigma^{(s)}_{R,{\rm IR}}
 =
 \frac{Q_S^2\alpha'}{12\pi}
 \ln\frac{q_{\rm cut}^2}{\mu^2}\,\sigma_0,
\end{equation}
which is finite in the limit $m_{A'}\to 0$. 
Therefore the dark-Higgs channel is IR safe at the inclusive level.

\paragraph{Dark-fermion channel.}
We now turn to the $\chi$ channel and directly adopt the limit $m_\chi\to 0$. 
The virtual correction is then governed by the small-$m_{A'}$ dependence of Eq.~(\ref{eq:3_15}), and the logarithmically singular part is
\begin{align} \label{eq:3_42}
 \sigma^{(\chi)}_{V,{\rm IR}}
 &=
 2\,\mathrm{Re}\,\mathcal C^{(\chi)}_{V,{\rm IR}}\,\sigma_0
 =
 \frac{\alpha'Q_\chi^2}{3\pi}
 \ln\frac{m_{A'}^2}{\mu^2}\,\sigma_0 .
\end{align}
For the real correction, the resonance-subtraction term ``$\sigma_0\, \mathrm{Br}(A'\to\chi\bar\chi)\, \mathcal{P}_{\rm res}(Q^2)$'' is localized around $Q^2\simeq m_{A'}^2$ and integrates to the finite constant $\sigma_0\,\mathrm{Br}(A'\to\chi\bar\chi)$. 
Therefore it does not modify the large-$Q^2$ asymptotics that are responsible for the IR logarithm. 
Hence the IR extraction can be performed directly from the first term in Eq.~(\ref{eq:3_31}). 
Set $m_\chi=0$, use Eq.~(\ref{eq:3_26}), and remove all the terms which are negligible when $m^2_{A'} \ll Q^2$.  To avoid integrating the widthless approximation through the subtracted pole, start the asymptotic integral at $Q^2=x_0m_{A'}^2$ with fixed $x_0>1$:
\begin{equation} \label{eq:3_43}
 \sigma^{(\chi)}_{R,{\rm IR}}=
 \frac{\alpha'Q_\chi^2}{3\pi}
 \int_{x_0m_{A'}^2}^{q_{\rm cut}^2} dQ^2\,
 \hat\sigma_0(s,Q^2)
 \frac{1}{Q^2} .
\end{equation}
Introducing the dimensionless variable
\begin{equation} \label{eq:3_44}
 x\equiv\frac{Q^2}{m_{A'}^2}\, ,
\end{equation}
Eq.~(\ref{eq:3_43}) becomes
\begin{equation} \label{eq:3_45}
 \sigma^{(\chi)}_{R,{\rm IR}} =
 \frac{\alpha'Q_\chi^2}{3\pi}
 \int_{x_0}^{x_{\rm max}} dx\,
 \hat\sigma_0(s,m_{A'}^2x)
 \frac{1}{x},
\end{equation}
with again $x_{\rm max}=q_{\rm cut}^2/m_{A'}^2$.  The fixed lower limit only changes finite constants:
\begin{equation} \label{eq:3_46}
\int_{x_0}^{x_{\rm max}}\frac{dx}{x}
=
\ln\frac{q_{\rm cut}^2}{m_{A'}^2}+{\cal O}(1)\, ,
\end{equation}
while
\begin{equation} \label{eq:3_47}
 \hat\sigma_0(s,m_{A'}^2x)=\sigma_0\left[1+{\cal O}\!\left(\frac{m_{A'}^2(x-1)}{s}\right)\right].
\end{equation}
The second term again yields only a finite contribution after integration, and therefore the real IR part is
\begin{equation} \label{eq:3_48}
 \sigma^{(\chi)}_{R,{\rm IR}} =
 \frac{\alpha'Q_\chi^2}{3\pi}\,\sigma_0
 \int_{x_0}^{x_{\rm max}} dx\,\frac{1}{x} =
 -\frac{\alpha'Q_\chi^2}{3\pi}
 \ln\frac{m_{A'}^2}{q_{\rm cut}^2}\,\sigma_0\, .
\end{equation}
Here we only retain the logarithmically singular term. 
Combining Eqs.~(\ref{eq:3_42}) and (\ref{eq:3_48}), we obtain
\begin{equation} \label{eq:3_49}
 \sigma^{(\chi)}_{V,{\rm IR}}+\sigma^{(\chi)}_{R,{\rm IR}} =
 \frac{\alpha'Q_\chi^2}{3\pi}
 \ln\frac{q_{\rm cut}^2}{\mu^2}\,\sigma_0\, ,
\end{equation}
which is finite in the limit $m_{A'}\to 0$. 
Therefore the dark-fermion channel is also IR safe at the inclusive level. 

\subsection{Numerical results \label{sec:Xsection_4}}

We now illustrate the different fixed-order components in the resonance-subtracted scheme. 
As a benchmark point we take
\begin{equation} \label{eq:3_50}
\sqrt{s}=10~\GeV \ , \ \mu=10~\GeV  \ , \  r=1 \ , \ 
 \alpha'=0.3 \ , \  Q_S=1 \ , \  Q_\chi=1  \ , \  m_\chi\to 0\, ,
\end{equation}
and we further set $E_\gamma^{\rm cut}=1~\GeV$ so that $q_{\rm cut}^2=s-2E_\gamma^{\rm cut}\sqrt{s}=80~\GeV^2$. 
For the invisible benchmark considered here we also take $\mathrm{Br}(A'\to\chi\bar\chi)=1$. 

\begin{figure}[t]
\centering
\includegraphics[width=0.75\textwidth]{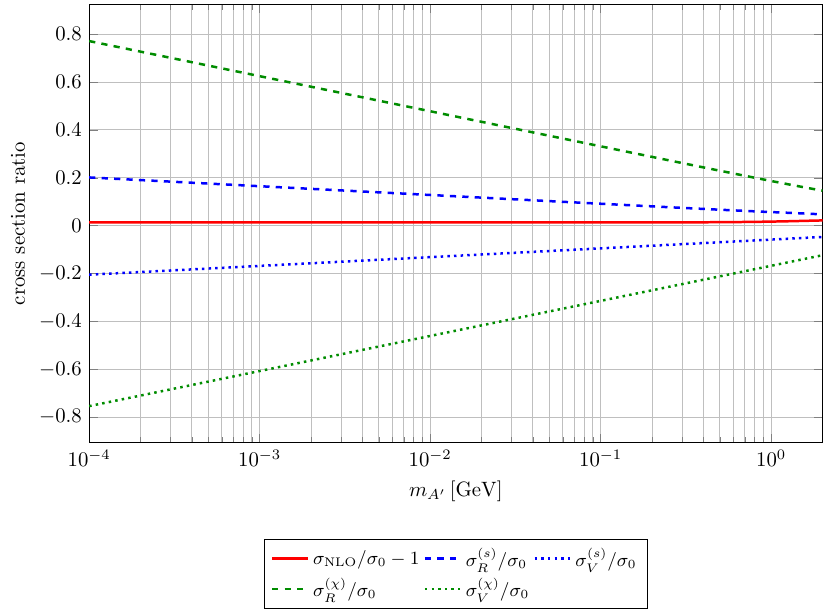}
\caption{Numerical illustration of the fixed-order NLO correction in the resonance-subtracted scheme. 
All curves are normalized to $\sigma_0$. 
The five curves correspond to $\sigma_{\rm NLO}/\sigma_0-1$ (red solid), $\sigma_R^{(s)}/\sigma_0$ (blue dashed), $\sigma_V^{(s)}/\sigma_0$ (blue dotted), $\sigma_R^{(\chi)}/\sigma_0$ (green dashed), and $\sigma_V^{(\chi)}/\sigma_0$ (green dotted). 
The benchmark parameters are $\sqrt{s}=10~\GeV$, $\mu=10~\GeV$, $r=1$, $\alpha'=0.3$, $Q_S=1$, $Q_\chi=1$, $m_\chi\to 0$, $E_\gamma^{\rm cut}=1~\GeV$, and $\mathrm{Br}(A'\to\chi\bar\chi)=1$.}
\label{fig:sec3ratio}
\end{figure}

The corresponding numerical result is shown in Fig.~\ref{fig:sec3ratio}. 
All five curves are normalized by the LO two-body cross section $\sigma_0$. 
The displayed curves are $\sigma_{\rm NLO}/\sigma_0-1$, $\sigma_R^{(s)}/\sigma_0$, $\sigma_V^{(s)}/\sigma_0$, $\sigma_R^{(\chi)}/\sigma_0$, and $\sigma_V^{(\chi)}/\sigma_0$. 
The red curve directly measures the net NLO correction. 
One can clearly see that both the $s$ and $\chi$ channels exhibit the expected real--virtual cancellation pattern discussed analytically above: the real contributions are positive, the virtual contributions are negative in the small-$m_{A'}$ region, and the full combination remains finite and smooth. 
As a result, the red curve remains moderate over the whole range $m_{A'}\in(10^{-4},2)~\GeV$. 
The net correction approaches $1.30\%$ at small masses and reaches $2.17\%$ at $m_{A'}=2~\GeV$.

\section{Fixed-order missing-mass spectrum and collinear enhancement \label{sec:wo_res}}

In this section we study the distribution of squared missing mass $M_X^2$ at NLO. 
The variable $M_X^2$, defined by $M^2_X \equiv s-2E_\gamma \sqrt{s}$, is the invariant mass squared of the invisible final-state system. 
At LO the invisible system consists of a single on-shell DP, so the partonic distribution would be a delta function at $M_X^2=m_{A'}^2$. 
However, such a strictly on-shell expression is not suitable for numerical illustration. 
For numerical illustration and consistency with the resonance-subtraction prescription, we represent the Born pole by the normalized Breit–Wigner profile \(\mathcal P_{\rm res}\)
\begin{equation} \label{eq:4_01}
 \frac{d\sigma_{\rm LO}}{dM_X^2} =
 \sigma_0\,\mathcal{P}_{\rm res}(M_X^2)\, ,
\end{equation}
with $\mathcal{P}_{\rm res}$ the normalized resonant distribution given in Eq.~(\ref{eq:3_27}):
\begin{equation} \label{eq:4_02}
 \mathcal{P}_{\rm res}(Q^2) =
 \frac{1}{\mathcal{N}_{\rm res}}
 \frac{1}{\pi}
 \frac{m_{A'}\Gamma_{A'}}{(Q^2-m_{A'}^2)^2+m_{A'}^2\Gamma_{A'}^2} \ , \  \mathcal{N}_{\rm res} = \int_{4m_\chi^2}^{q_{\rm cut}^2} dQ^2\,
 \frac{1}{\pi}
 \frac{m_{A'}\Gamma_{A'}}{(Q^2-m_{A'}^2)^2+m_{A'}^2\Gamma_{A'}^2}
\end{equation}
By construction, this profile integrates to unity over the physical range $4m_\chi^2 <M_X^2<q_{\rm cut}^2$, and peaks around $m^2_{A'}$ with width $\Gamma_{A'}$.

The NLO distribution, which already includes the resonant LO distribution, can be decomposed as
\begin{equation} \label{eq:4_03}
 \frac{d\sigma_{\rm NLO}}{dM_X^2} = 
 \left(\sigma_0+\sigma_V^{(s)}+\sigma_V^{(\chi)}\right)\mathcal{P}_{\rm res}(M_X^2)
 + \frac{d\sigma_R^{(s)}}{dM_X^2}
 + \frac{d\sigma_R^{(\chi)}}{dM_X^2}\, ,
\end{equation}
with 
\begin{equation} \label{eq:4_04}
 \frac{d\sigma_R^{(s)}}{dM_X^2} =
 \hat\sigma_0(s,M_X^2)
 \frac{Q_S^2\alpha' m_{A'}^2}{\pi M_X^2(M_X^2-m_{A'}^2)^2}
 \lambda^{1/2}(M_X^2,m_{A'}^2,m_s^2)
 \left[ 1+ \frac{\lambda(M_X^2,m_{A'}^2,m_s^2)}{12m_{A'}^2M_X^2} \right]\, ,
\end{equation}
and
\begin{equation} \label{eq:4_05}
 \frac{d\sigma_R^{(\chi)}}{dM_X^2} =
 \hat\sigma_0(s,M_X^2)
 \frac{1}{\pi}
 \frac{M_X\Gamma_\chi(M_X^2)}{(M_X^2-m_{A'}^2)^2+m_{A'}^2\Gamma_{A'}^2}
 -  \sigma_0\, \mathcal{P}_{\rm res}(M_X^2)\, .
\end{equation}
In the last line we already set $\mathrm{Br}(A'\to\chi\bar\chi) = 1$. 
And it is clear that the sum of $d \sigma_R^{(\chi)}$ and $d \sigma_{\rm LO}$ gives the complete width-regulated tree-level three-body distribution for the $\gamma \chi \bar{\chi}$ final state. 

\subsection{Numerical results \label{sec:wo_res_numerical}}

\begin{figure}[!htbp]
\centering
\includegraphics[width=0.98\textwidth]{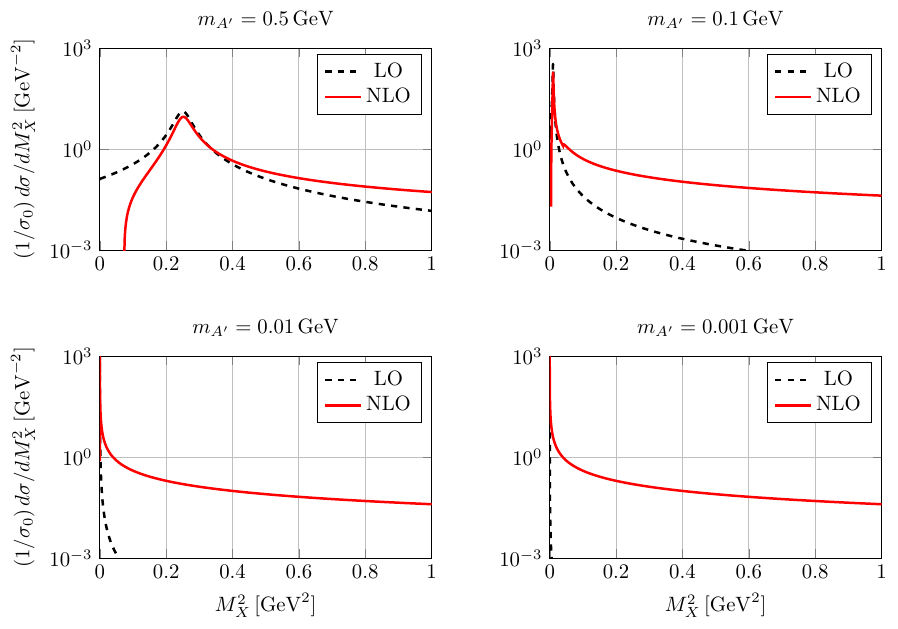}
\caption{LO and NLO $M_X^2$ distributions without resummation. 
All curves are normalized to $\sigma_0$. 
The dashed black line denotes $d\sigma_{\rm LO}/dM_X^2$, while the red solid line denotes the full fixed-order result $d\sigma_{\rm NLO}/dM_X^2$. 
The benchmark parameters are $\sqrt{s}=10~\GeV$, $\mu=10~\GeV$, $r=1$, $\alpha'=0.3$, $Q_S=1$, $Q_\chi=1$, $m_\chi\to 0$, $E_\gamma^{\rm cut}=1~\GeV$, and $\mathrm{Br}(A'\to\chi\bar\chi)=1$. 
The four panels correspond to $m_{A'}=0.5~\GeV$, $0.1~\GeV$, $0.01~\GeV$, and $0.001~\GeV$.}
\label{fig:M2X_nores}
\end{figure}

In Fig.~\ref{fig:M2X_nores} we present the LO and NLO $M_X^2$ distributions with the following setting
\begin{equation}\label{eq:4_06}
\begin{gathered}
 \sqrt{s}=10~\GeV\ ,\quad\mu=10~\GeV\ ,\quad r=1\ ,\quad\alpha'=0.3\ ,\\
 Q_S=1\ ,\quad Q_\chi=1\ ,\quad m_\chi\to0\ ,\quad E_\gamma^{\rm cut}=1~\GeV\ ,
\end{gathered}
\end{equation}
and $m_{A'}$ takes 4 values: 0.5 GeV,  0.1 GeV, 0.01 GeV, and 0.001 GeV. 
The dashed black line denotes the width-smeared LO distribution and the red solid line denotes the full NLO result $d\sigma_{\rm NLO}/dM_X^2$. 
For $m_{A'}=0.5~\GeV$, both distributions exhibit a pronounced resonance near $M_X^2=m_{A'}^2$.
As $m_{A'}$ and $m_s$ decrease at fixed $r=1$, the resonance moves towards the left edge and becomes increasingly narrow on the common linear horizontal scale, while the NLO real-emission continuum becomes more prominent relative to the LO tail.
The hierarchy $m_{A'}^2,m_s^2\ll M_X^2\ll s$ then extends to progressively smaller $M_X^2$, revealing the $1/M_X^2$ behavior derived below over a wider range.
The resulting rise towards the small-$M_X^2$ endpoint illustrates how the masses regulate the collinear singularity and how the singular behavior emerges as the dark masses decrease, even though the inclusive NLO rate remains finite.

\subsection{Analytic origin of the collinear enhancement \label{sec:wo_res_analytic}}

The behavior seen in the numerical distribution can be understood analytically.  
Consider the hierarchy
\begin{equation} \label{eq:4_08a}
 m_{A'}^2,\ m_s^2,\ m_\chi^2 \ll M_X^2 \ll s,
\end{equation}
and stay away from the narrow resonant region $M_X^2\simeq m_{A'}^2$.  
In this region the off-shell two-body cross section is smooth,
\begin{equation} \label{eq:4_08b}
 \hat\sigma_0(s,M_X^2)=\sigma_0\left[1+{\cal O}\left(\frac{M_X^2-m_{A'}^2}{s}\right)\right],
\end{equation}
where the difference between $\hat\sigma_0(s,M_X^2)$ and $\sigma_0$ only affects nonsingular terms in the small-mass expansion.

For the dark-Higgs channel, the $\lambda$ function satisfies
\begin{equation} \label{eq:4_08c}
 \lambda^{1/2}(M_X^2,m_{A'}^2,m_s^2)=M_X^2+{\cal O}(m_{A'}^2,m_s^2),
\end{equation}
and the term proportional to $\frac{\lambda}{12m_{A'}^2M_X^2}$ in Eq.~(\ref{eq:4_04}) is the leading contribution in the light-mass limit.  
And thus $d\sigma_R^{(s)}$ can be expanded as 
\begin{equation} \label{eq:4_08d}
 \frac{1}{\sigma_0}\frac{d\sigma_R^{(s)}}{dM_X^2} = 
 \frac{\alpha' Q_S^2}{12\pi}\,\frac{1}{M_X^2}\left[1 + {\cal O}\!\left(\frac{m_{A'}^2}{M_X^2},\frac{m_s^2}{M_X^2},\frac{M_X^2}{s}\right)\right]\, .
\end{equation}
The dark-Higgs real contribution therefore has a $1/M_X^2$ enhancement in the hierarchy of Eq.~(\ref{eq:4_08a}). This expression is an asymptotic continuum result, not an extrapolation through the finite-mass threshold or the resonant region.

A similar analysis can be performed on the dark fermion channel as well. 
Firstly, in the $m_\chi\to0$ limit one has
\begin{equation} \label{eq:4_08e}
 \Gamma_\chi(M_X^2)=\frac{\alpha'Q_\chi^2}{3}\,M_X\, .
\end{equation}
Thus away from the subtracted resonance, Eq.~(\ref{eq:4_05}) then gives
\begin{equation} \label{eq:4_08f}
 \frac{1}{\sigma_0}\frac{d\sigma_R^{(\chi)}}{dM_X^2} = 
 \frac{\alpha' Q_\chi^2}{3\pi}\,\frac{1}{M_X^2} \left[1+  {\cal O}\!\left(\frac{m_{A'}^2}{M_X^2},\frac{m_\chi^2}{M_X^2},\frac{M_X^2}{s}\right)\right]\, .
\end{equation}

In the light-mass hierarchy, small $M_X^2$ relative to the hard scale corresponds to nearly collinear daughters of the intermediate $A'$. Combining Eqs.~(\ref{eq:4_08d}) and (\ref{eq:4_08f}) gives the leading continuum behavior
\begin{equation} \label{eq:4_08g}
 \frac{1}{\sigma_0}\frac{d \sigma_{\rm NLO}}{dM_X^2} \bigg|_{\rm col} =
 \frac{\alpha'}{\pi}\left(
 \frac{Q_S^2}{12}+\frac{Q_\chi^2}{3}
 \right)\frac{1}{M_X^2}\, .
\end{equation}
Its integral gives the logarithm $\ln({q_{\rm cut}^2}/{m_{\rm dark}^2})$ ($m_{\rm dark}$ is the mass scale of $A'$ and $s$) that cancels against the corresponding virtual correction in the inclusive rate.  
However, the cancellation is not local in $M_X^2$: the virtual term is tied to the width-smeared two-body line shape, whereas the real-emission contribution populates a continuous invariant-mass tail.  
This is why the inclusive total NLO cross section can be finite while the NLO differential distribution still displays a collinear enhancement.  
In the next section we resum the enhanced logarithms in this hierarchy. 

\section{Resummation of the primary and secondary radiation\label{sec:resum}}

We now construct the resummed missing-mass spectrum analytically, starting from the splitting kernels extracted from the amplitudes. 
The primary evolution and the coherent secondary contribution are organized explicitly, so that their dependence on the dark charges can be followed without generating MC events. 
We use Sudakov exponentiation~\cite{Sudakov:1954sw} and event-shape resummation~\cite{Dasgupta:2003iq,Banfi:2004yd,Banfi:2018mcq}, together with factorization and evolution methods for hard QCD processes and parton densities~\cite{Collins:1989gx,Altarelli:1977zs,Curci:1980uw}.

The fixed-order analysis in Sec.~\ref{sec:wo_res_analytic} shows that the small-$M_X^2$ distribution contains the quasi-collinear singular term when the daughter momenta become nearly parallel:
\begin{equation} \label{eq:5_00}
 {1\over \sigma_0}{d\sigma_{\rm NLO} \over dM_X^2} \simeq
 \frac{\alpha'}{\pi}\left(
 \frac{Q_S^2}{12}+\frac{Q_\chi^2}{3}
 \right)\frac{1}{M_X^2}\, .
\end{equation}
For later convenience we define the primary radiation coefficient 
\begin{equation} \label{eq:5_00a}
 a_{\rm pri} \equiv  \frac{\alpha'}{\pi}\left(
 \frac{Q_S^2}{12}+\frac{Q_\chi^2}{3}
 \right)  \, ,
\end{equation}
and thus in the collinear region one has a concise expression
\begin{equation} \label{eq:5_00b}
 {1\over \sigma_0}{d\sigma_{\rm NLO} \over dM_X^2} \simeq \frac{ a_{\rm pri}}{M_X^2}\, .
\end{equation}

In this section we concentrate on the following mass hierarchy
\begin{equation} \label{eq:5_01}
 m_{A'}^2,\ m_s^2,\ m_\chi^2 \ll M_X^2\ll Q^2_{\rm hard} \sim s ,
\end{equation}
so that all dark sector particles can be treated as massless in our analysis.
Here $Q_{\rm hard}$ is the hard scale of the invisible branch, which is of the same order of magnitude as the center-of-mass energy.
We use a fixed-coupling Sudakov treatment and retain the leading-power terms in this hierarchy. Endpoint limits below refer to this light-mass approximation; they are not finite-mass predictions arbitrarily close to threshold. 

\subsection{Primary Sudakov factor \label{sec:resum_primary}}

As we discussed in Sec.~\ref{sec:intro}, the evolution process of the final state DP can be called dark FSR (final state radiation), and the differential cross-section can be factorized as
\begin{equation} \label{eq:5_02a}
d \sigma_{\gamma+X} \simeq d \sigma_{\gamma + {A'}^\ast}\times d {\cal P}_{{A'}^\ast \to X}
\end{equation}
Actually, this factorization has been implemented in Eq.~(\ref{eq:3_22}) and Eq.~(\ref{eq:3_25}). 

At leading power, the process $e^+e^-\to \gamma A'$ produces a transversely polarized DP.  
From now on, we will use $A'_T$ and $A'_L$ (or simply $T$ and $L$) to label the transverse DP and the longitudinal DP, respectively. 
In the massless and collinear limit, the radiation/splitting of $A'_T$ is dominated by~\cite{Li:2025tlg}
\begin{equation} \label{eq:5_03}
 A'_T\to A'_L+s \ , \ 
 A'_T\to\chi\bar\chi \, ,
\end{equation}
and the differential splitting probability can be written as 
\begin{equation} \label{eq:5_04}
 d{\cal P}^{(c)}_{\rm pri} = {\alpha'\over 2\pi}\,C_{\rm pri}^{(c)}\, P^{(c)}_{\rm pri}(z)\,dz\, {dt \over t}\, , 
\end{equation}
where $t$ is the virtuality generated by the primary splitting.
In the massless and collinear region, $t$ is identified with $M_X^2$ in the singular approximation.  
The label $c=S,F$ is the tag for the scalar/fermion channel.
In Eq.~(\ref{eq:5_04}), $C_{\rm pri}^{(S)}=Q^2_S$ and $C_{\rm pri}^{(F)}=Q^2_\chi$.  
$z$ is the energy fraction of the first listed daughter and $\bar z\equiv 1-z$ is the energy fraction of the second daughter.
And the splitting kernels in Eq.~(\ref{eq:5_04}) are~\cite{Li:2025tlg}
\begin{equation} \label{eq:5_05}
P^{(S)}_{\rm pri}(z) \equiv P_{T\to Ls}(z)=z\bar z \ , \
P^{(F)}_{\rm pri}(z) \equiv P_{T\to \chi\bar\chi}(z)=z^2+\bar z^2\, .
\end{equation}

Since
\begin{equation} \label{eq:5_06}
 \int_0^1 dz\,P_{T\to Ls}(z)={1\over6}\ , \ 
 \int_0^1 dz\,P_{T\to\chi\bar\chi}(z)={2\over3}\, ,
\end{equation}
Eq.~(\ref{eq:5_04}) reproduces the NLO singular coefficient in Eq.~(\ref{eq:5_00}) after integrating over $z$, i.e.,
\begin{equation} \label{eq:5_07}
 d{\cal P}_{T\to X}^{\rm pri} =\sum_c \int_0^1 dz\,  \frac{d{\cal P}^{(c)}_{\rm pri}}{dz} = {\alpha'\over 2\pi}\, \left(Q_S^2\frac{1}{6} +  Q_\chi^2\frac{2}{3} \right)\, \frac{d t}{t} = a_{\rm pri} \frac{d t}{t}\, ,
\end{equation}
with the factor $a_{\rm pri}$ defined in Eq.~(\ref{eq:5_00a}).
It should be emphasized that the kernels in Eq.~(\ref{eq:5_05}) are finite at both $z=0$ and $z=1$.  
Therefore the primary splitting contains the collinear measure $dt/t$, but it does not contain a soft endpoint pole.  

The Sudakov factor is actually the probability of no-radiation between the hard scale $Q^2_{\rm hard}$ and the measured scale $M^2_X$. 
For the primary radiation process, the corresponding Sudakov factor is given by
\begin{equation} \label{eq:5_08}
 \Delta_{\rm pri}(Q^2_{\rm hard},M_X^2) = \exp[ - \int^{Q^2_{\rm hard}}_{M_X^2}  d{\cal P}_{T\to X}^{\rm pri}  ]
 =  \exp[-a_{\rm pri} \ln {Q_{\rm hard}^2\over M_X^2} ]\, . 
\end{equation}
Then, to calculate the primary-resummed $M^2_X$ distribution, the singular NLO expression Eq.~(\ref{eq:5_00b}) is multiplied by the Sudakov factor $\Delta_{\rm pri}(Q^2_{\rm hard},M_X^2)$:
\begin{equation} \label{eq:5_09}
{1\over \sigma_0}{d\sigma_{\rm resum}^{\rm pri} \over dM_X^2}  = \Delta_{\rm pri}(Q^2_{\rm hard},M_X^2) \times {1\over \sigma_0}{d\sigma_{\rm NLO} \over dM_X^2}\bigg|_{\rm col} = \frac{ a_{\rm pri}}{M_X^2} \exp[-a_{\rm pri} \ln {Q_{\rm hard}^2\over M_X^2} ] \, . 
\end{equation}
The Sudakov factor accounts for the veto on resolvable primary splittings above the measured scale. Multiplication by the splitting density gives the corresponding primary endpoint distribution. 

Equation~(\ref{eq:5_09}) changes the $M_X^2$ dependence from the non-integrable fixed-order form $(M_X^2)^{-1}$ to $(M_X^2)^{-1+a_{\rm pri}}$. 
For $0<a_{\rm pri}<1$ the differential density is still enhanced pointwise as $M_X^2\to0$, but it is integrable:
\begin{equation} \label{eq:5_09_integral}
\int_0^{M_{X,0}^2}dM_X^2\,{a_{\rm pri}\over M_X^2}\left({M_X^2\over Q_{\rm hard}^2}\right)^{a_{\rm pri}}
=\left({M_{X,0}^2\over Q_{\rm hard}^2}\right)^{a_{\rm pri}}<\infty\, .
\end{equation}
The primary Sudakov factor therefore gives an integrable endpoint density whose formal massless cumulative distribution is normalized at $M_X^2=Q_{\rm hard}^2$. 

\subsection{Secondary Sudakov factor \label{sec:resum_secondary}}

We now consider soft radiation from the daughter system produced by the primary splitting.
For the dark fermion channel, the two secondary splittings are
\begin{equation} \label{eq:5_10}
 \chi\to\chi A'_T \ , \
 \bar\chi\to\bar\chi A'_T\, .
\end{equation}
For the dark Higgs channel, the high-energy Goldstone-equivalence limit treats $A'_L$ and $s$ as the two real components of the charged dark Higgs field.  
The corresponding splitting processes are
\begin{equation} \label{eq:5_11}
 A'_L\to A'_T s\ , \
 s\to A'_T A'_L\, .
\end{equation}
The splitting kernels for the above processes are
\begin{equation} \label{eq:5_12}
P_F(x) = {1+(1-x)^2\over x}\ , \
P_S(x) = {2(1-x)\over x}\, .
\end{equation}
with $x$ chosen as the energy fraction of the emitted transverse DP. 
$P_F(x)$ applies to $\chi\to\chi A'_T$ and $\bar\chi\to\bar\chi A'_T$, while $P_S(x)$ is the scalar-QED kernel when $x$ is the transverse-DP energy fraction.

Unlike the primary kernels in Eq.~(\ref{eq:5_05}), these local collinear kernels have soft poles.
Thus at first sight, the secondary splitting would provide the double-log suppression usually seen in QCD processes. 
But the parent Abelian $A'_T$ carries no $U(1)'$ charge; the daughters (which form a $U(1)'$ dipole) cannot in general be treated as independent soft sources before their interference is included.  
Next we perform a detailed analysis. 

\begin{figure}[!htbp]
\centering
\includegraphics[width=0.4\linewidth]{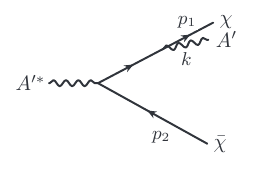}
\caption{Schematic illustration of the dipole radiation in the dark-fermion channel.
The virtual dark photon $A^{\prime *}$ first splits into a $\chi\bar{\chi}$ pair,
and the secondary dark photon with momentum $k$ can be emitted from either daughter.
The daughter momenta are denoted by $p_1$ and $p_2$.}
\label{fig:sec5_dipole}
\end{figure}

For the fermion channel the soft limit can be established at the amplitude level.  The soft current is
\begin{equation} \label{eq:5_12a}
J^\mu(k)=Q_\chi\left({p_1^\mu\over p_1\cdot k}-{p_2^\mu\over p_2\cdot k}\right)\, ,
\qquad k_\mu J^\mu=0\, ,
\end{equation}
with $k$, $p_1$, and $p_2$ the momenta for the secondary $A'_T$ (radiated from $\chi$ or $\bar{\chi}$), $\chi$ (1st daughter), and $\bar{\chi}$ (2nd daughter), respectively. 
See Fig.~\ref{fig:sec5_dipole} for illustration. 

For massless daughters, the cross-section level squared current is 
\begin{equation} \label{eq:5_12b}
-J^2(k)={2Q_\chi^2\,p_1\cdot p_2\over(p_1\cdot k)(p_2\cdot k)}\, ,
\end{equation}
and the corresponding dipole differential probability is given by
\begin{equation} \label{eq:5_12b1}
d{\cal P}_{\rm dipole}={\alpha'\over4\pi^2}{d^3k\over k^0}[-J^2(k)]\, .
\end{equation}

Thus the angular antenna is proportional to
\begin{equation} \label{eq:5_12c}
{1-\cos\theta\over(1-\cos\theta_{1})(1-\cos\theta_{2})}\, ,
\end{equation}
where $\theta=\angle(p_1,p_2)$ denotes the opening angle of the primary splitting, 
while $\theta_1=\angle(p_1,k)$ and $\theta_2=\angle(p_2,k)$ are the angles between the secondary radiation and the two charged daughters, respectively.
For a secondary emission at an angle much larger than $\theta$ ($\theta_1\sim \theta_2 \gg \theta$), the two terms in Eq.~(\ref{eq:5_12a}) cancel at leading power ($d{\cal P}_{\rm dipole} \propto d \theta_i^2 /\theta_i^4 $).  
Wide-angle radiation is therefore suppressed by the small opening angle of the neutral dipole.
At leading-log accuracy the same cancellation can be represented probabilistically by angular ordering, in which an emission assigned to daughter $i$ obeys $\theta_i<\theta$.

In the massless limit, the virtuality from the primary splitting is
\begin{equation} \label{eq:5_12d}
t_{\rm pri}\simeq  z\bar z Q_{\rm hard}^2 \theta^2\, .
\end{equation}
After the secondary splitting, the total virtuality, which is actually the invariant mass squared of the dark branch, receives the contribution 
\begin{equation} \label{eq:5_13}
M_X^2 \simeq t_{\rm pri} + \sum_i \frac{t_{{\rm sec},i}}{\zeta_i} = t_{\rm pri} + \sum_i  \zeta_i Q_{\rm hard}^2x_i(1-x_i)\theta_{i}^2 \, .
\end{equation}
with $\zeta_1=z$, $\zeta_2=\bar z$ (the primary energy fractions), and $x_i$ the transverse-DP energy fraction relative to daughter $i$.

Now we need to calculate the secondary no-emission probability that vetoes all events with $\frac{t_{\rm sec,i}}{\zeta_i} > M^2_X - t_{\rm pri}$; in other words, the secondary splitting cannot cause the virtuality of the dark branch to exceed $M^2_X$~\footnote{Strictly speaking we should veto events with $\sum_i \frac{t_{\rm sec,i}}{\zeta_i} > M^2_X - t_{\rm pri}$. However, in the leading-log approximation we can simplify our calculation by replacing $\sum_i \frac{t_{\rm sec,i}}{\zeta_i} > M^2_X - t_{\rm pri}$ with $\frac{t_{\rm sec,i}}{\zeta_i} > M^2_X - t_{\rm pri}$.}.
The corresponding secondary Sudakov factor is
\begin{equation}\label{eq:5_14a}
\begin{aligned}
 &\Delta^{(c)}_{\rm sec}(Q^2_{\rm hard},M_X^2-t_{\rm pri},\theta)\\
 &\quad=\exp\Bigg[-\sum_{i=1}^{2}\int_0^1 dx_i\int_0^{\theta^2} \frac{d\theta_i^2}{\theta_i^2}
 \frac{\alpha'}{2\pi}\,C_i^{(c)}P_i^{(c)}(x_i)\,
 \Theta\!\left(\frac{t_{\rm sec,i}}{\zeta_i}-M^2_X+t_{\rm pri}\right)\Bigg].
\end{aligned}
\end{equation}
with $\Theta$ the Heaviside step function. 
$c=F,S$ labels the dark-fermion or dark-Higgs primary channels.
For the fermion channel,
\begin{equation}
 P_i^{(F)}(x)=P_F(x)\ , \ C_i^{(F)}=Q_\chi^2\, ,
\end{equation}
whereas for the dark-Higgs channel,
\begin{equation}
 P_i^{(S)}(x)=P_S(x) \ , \  C_i^{(S)}=Q_S^2\, .
\end{equation}

Let us first perform the integration with respect to $\theta_i^2$ in Eq.~(\ref{eq:5_14a}). 
The upper bound of $\theta_i$, as we discussed before, should be the primary splitting angle $\theta$. 
The lower bound of $\theta_i$ is determined by the Heaviside step function in Eq.~(\ref{eq:5_14a}):
\begin{equation}
\theta^2_{i,{\rm min}} =  \frac{M^2_X - t_{\rm pri}}{\zeta_i Q^2_{\rm hard}\, x_i(1-x_i)}
\end{equation}
Thus the integration over $\theta_i^2$ gives
\begin{equation}\label{eq:5_16}
\begin{aligned}
 &\Delta^{(c)}_{\rm sec}(Q^2_{\rm hard},M_X^2-t_{\rm pri},\theta)\\
 &\quad=\exp\Bigg[-\sum_{i=1}^{2}\int dx_i\,
 \ln\frac{\theta^2\zeta_i Q^2_{\rm hard}\,x_i(1-x_i)}{(M^2_X-t_{\rm pri})}
 \frac{\alpha'}{2\pi}\,C_i^{(c)}P_i^{(c)}(x_i)\\
 &\hspace{7em}\times\Theta\!\left(\theta^2\zeta_i Q^2_{\rm hard}\,x_i(1-x_i)-M^2_X+t_{\rm pri}\right)\Bigg].
\end{aligned}
\end{equation}

Next we perform the integration over $x_i$. 
It is useful to introduce 
\begin{equation}
\rho_i\equiv { M^2_X - t_{\rm pri} \over \theta^2 \zeta_i Q_{\rm hard}^2}
\end{equation}
Then the integral boundary for $x$ is given by 
\begin{equation}
 x_{i,-}<x<x_{i,+} \ , \
 x_{i,\pm}=\frac{1}{2}\left(1\pm\sqrt{1-4 \rho_i }\right)\, .
\end{equation}
And it is evident that the integral range of $x$ is non-zero only when
\begin{equation}
\rho_i =  \frac{M_X^2-t_{\rm pri}}{\theta^2\zeta_i Q_{\rm hard}^2}  < \frac{1}{4}\, .
\end{equation}
Thus Eq.~(\ref{eq:5_16}) can be rewritten as:
\begin{align} \label{eq:5_17}
&\Delta^{(c)}_{\rm sec}(Q^2_{\rm hard},M_X^2-t_{\rm pri},\theta) = \nonumber \\
&\ \ \ \ \ \ \ \ \   \ \ \ \ \   \exp[ - \sum_{i=1}^{2} \frac{\alpha'}{2\pi}\,C_i^{(c)} \Theta\!\left( \frac{1}{4} - \rho_i  \right)\, \int^{x_{i,+}}_{x_{i,-}} dx_i\, 
\ln \frac{  x_i(1-x_i)}{\rho_i} \,
P_i^{(c)}(x_i)\,   ]\, . 
\end{align}

\subsection{Endpoint scaling after primary--secondary convolution \label{sec:resum_num}}

We now determine how secondary radiation modifies the integrable primary endpoint spectrum. The distinction between a probability density and a cumulative probability is essential: the primary splitting density is convoluted with the secondary probability that its contribution remains below the available mass budget. The final mass distribution is obtained only after differentiating the combined cumulative expression. With primary splitting alone, the Sudakov factor gives
\begin{equation} \label{eq:5_21}
{\rm Prob}_{\rm pri}(< M^2_X) = \Delta_{\rm pri}(Q^2_{\rm hard},M^2_X) = \exp[-a_{\rm pri} \ln {Q_{\rm hard}^2\over M_X^2} ]
\end{equation}
The resummed $M^2_X$ distribution follows by differentiating ${\rm Prob}_{\rm pri}(< M^2_X)$:
\begin{equation} \label{eq:5_22}
  {1\over \sigma_0}{d\sigma_{\rm resum}^{\rm pri} \over dM_X^2}
  = \frac{d}{d M^2_X}  {\rm Prob}_{\rm pri}(< M^2_X) 
  = \frac{ a_{\rm pri}}{M_X^2} \exp[-a_{\rm pri} \ln {Q_{\rm hard}^2\over M_X^2} ] \, .
\end{equation}
This expression is consistent with Eq.~(\ref{eq:5_09}). 
It obeys $ {\rm Prob}_{\rm pri}(<Q_{\rm hard}^2)=1$ and Eq.~(\ref{eq:5_09_integral}), so every finite endpoint bin is finite.  

Considering the secondary splitting process, the ``primary + secondary'' cumulative distribution can be written in the convolution form
\begin{align} \label{eq:5_23}
&{\rm Prob}_{\rm pri+sec}(< M^2_X) = \\\nonumber
& \ \ \ \  \sum_c 
\int_0^{M^2_X} \int \underbrace{dz {d t_{\rm pri} \over t_{\rm pri} }\ \Delta_{\rm pri}(Q^2_{\rm hard},t_{\rm pri}) 
\times \left[ \frac{\alpha'}{2\pi} C^{(c)}_{\rm pri} P^{(c)}_{\rm pri}(z) \right]}_{\text{the differential probability for primary splitting at $z,t_{\rm pri}$}}
\times \Delta^{(c)}_{\rm sec}(Q^2_{\rm hard}, M^2_X - t_{\rm pri},\theta) 
\end{align}
Its differential distribution is
\begin{equation} \label{eq:5_24}
  {1\over \sigma_0}{d\sigma_{\rm resum}^{\rm pri+sec} \over dM_X^2}
  = \frac{d}{d M^2_X}  {\rm Prob}_{\rm pri+sec}(< M^2_X) \, .
\end{equation}

The dependence of the secondary Sudakov factor on the primary angle is important for the endpoint behavior.  
Using Eq.~(\ref{eq:5_12d}) in the definition of $\rho_i$, and introducing $y=t_{\rm pri}/M_X^2$, we find
\begin{equation} \label{eq:5_25}
\rho_1={M_X^2-t_{\rm pri}\over t_{\rm pri}}\bar z={1-y\over y}\bar z\, ,\qquad
\rho_2={M_X^2-t_{\rm pri}\over t_{\rm pri}}z={1-y\over y}z\, .
\end{equation}
In particular, $Q_{\rm hard}$ cancels from both secondary logarithms.  It is useful to denote the complete $x_i$ integral in Eq.~(\ref{eq:5_17}) by
\begin{equation}\label{eq:5_26}
\begin{aligned}
 {\cal D}_c(y,z)&=\exp\left\{-{\alpha'Q_c^2\over2\pi}\left[
 G_c\left({1-y\over y}\bar z\right)+G_c\left({1-y\over y}z\right)\right]\right\}\, ,\\
 {\rm with }\quad G_c(\rho)&=\Theta\left({1\over4}-\rho\right)
 \int_{x_-}^{x_+}dx\,P_c(x)\ln{x(1-x)\over\rho}\, ,\\
 {\rm and }\quad Q_c&=\left\{Q_\chi\ (c=F),\,Q_S\ (c=S)\right\}\, .
\end{aligned}
\end{equation}
Substituting it into Eq.~(\ref{eq:5_23}), the dependence on $M_X^2$ can be separated:
\begin{equation} \label{eq:5_27}
{\rm Prob}_{\rm pri+sec}(<M_X^2) =
 {\cal C}_{\rm sec} \left({M_X^2\over Q_{\rm hard}^2}\right)^{a_{\rm pri}}
=  {\cal C}_{\rm sec}\times {\rm Prob}_{\rm pri}(<M_X^2)\, ,
\end{equation}
with the secondary factor ${\cal C}_{\rm sec}$ given by 
\begin{equation} \label{eq:5_28}
{\cal C}_{\rm sec} = \sum_c{\alpha'C_{\rm pri}^{(c)}\over2\pi}
\int_0^1dz\,P_{\rm pri}^{(c)}(z)
\int_0^1dy\,y^{a_{\rm pri}-1}{\cal D}_c(y,z)\, .
\end{equation}
Within the fixed-coupling, leading-power, angular-ordered approximation, ${\cal C}_{\rm sec}$ is dimensionless and independent of $M_X^2$. 
Consequently, the secondary contribution does not introduce an additional double-logarithmic exponent in the measured mass, and the endpoint power in Eq.~(\ref{eq:5_27}) remains controlled by $a_{\rm pri}$. 

We evaluate Eq.~(\ref{eq:5_28}) numerically. Table~\ref{tab:sec5_cnum} shows that the deviation from unity remains below \(2\%\) for different $\alpha'$. 
Thus the endpoint distribution is mainly controlled by the primary resummation. 

\begin{table}[h]
\centering
\caption{Numerical values of ${\cal C}_{\rm sec}$ in Eq.~(\ref{eq:5_28}) for $Q_\chi=Q_S=1$ with different $\alpha'$. }
\label{tab:sec5_cnum}
\begin{tabular}{cc}
\hline\hline
$\alpha'$ & ${\cal C}_{\rm sec}$ \\
\hline
0.1 & 0.998840 \\
0.3 & 0.992006 \\
0.5 & 0.981812 \\
\hline\hline
\end{tabular}
\end{table}

\section{Signal shape and BaBar limit reinterpretation \label{sec:real_collider}}

We now use the particle-level calculation we performed in the last several sections to find the signal shape in a monophoton search.
The BaBar experiment serves as an illustrative benchmark.
BaBar searched for $e^+e^-\to\gamma A'$ followed by an invisible decay using $53~\mathrm{fb}^{-1}$ of data and reported limits for $m_{A'}\leq 8~\GeV$~\cite{BaBar:2017tiz}.  
The experimentally reconstructed variable is
\begin{equation} \label{eq:6_mx_definition}
M_{X,{\rm obs}}^2=s-2E_{\gamma,{\rm obs}}\sqrt{s}\, .
\end{equation}
At particle level, $M_X^2$ is the invariant mass squared of the invisible dark branch.  
When $m_s, m_{A'} \sim \sqrt{s}$, the fixed order results given in Sec.~\ref{sec:Xsection} and Sec.~\ref{sec:wo_res} are enough to obtain a physical distribution of $M_X^2$. 
But for $m_s, m_{A'} \ll \sqrt{s}$, we need to use a ``matching'' method to combine the resummed result given in Sec.~\ref{sec:resum} with the fixed order distribution. 

For convenience, in this section we define \(
q\equiv Q^2\equiv M_X^2\) and \(q_{\max}\equiv q_{\rm cut}^2\).

\subsection{Phase space cuts \label{sec:phase_space_cuts}}

In this subsection we discuss how the phase-space cuts used in the BaBar analysis affect the total rate and distribution. 
BaBar divides the accepted data into LowM and HighM selections.  
We focus on the LowM selection and impose the corresponding phase-space cut
\begin{equation} \label{eq:6_lowm_cuts}
E_\gamma>3~\GeV\, ,\qquad |\cos\theta_\gamma|<0.6\, .
\end{equation}
The photon-energy cut fixes the upper limit $q_{\rm max}=s-2E_\gamma^{\rm cut}\sqrt{s}$.  
The angular cut removes the forward and backward collinear regions, so the production kernel can be evaluated with $m_e=0$.  
Defining $c\equiv\cos\theta_\gamma$ and $\rho\equiv q/s$, the spin-averaged squared matrix element for $e^+e^-\to\gamma {A'}^*(q)$ is
\begin{equation} \label{eq:6_amplitude_squared}
\overline{|{\cal M}|^2}
=2\varepsilon^2e^4
\left({u\over t}+{t\over u}+{2qs\over tu}\right)
=4\varepsilon^2e^4
\left[{1+c^2\over1-c^2}
+{4\rho\over(1-\rho)^2(1-c^2)}\right] .
\end{equation}
Here $t$ and $u$ are Mandelstam variables. 
The differential production kernel is therefore
\begin{equation} \label{eq:6_angular_kernel}
{d\hat\sigma_0(s,q)\over dc}
={2\pi\varepsilon^2\alpha^2\over s}(1-\rho)
\left[{1+c^2\over1-c^2}+{4\rho\over(1-\rho)^2(1-c^2)}\right]  \, .
\end{equation}
After integrating over the accepted angular range, $-c_{\rm max}<c<c_{\rm max}$ with $c_{\rm max}=0.6$, we obtain
\begin{equation} \label{eq:6_angular_integrated}
\hat\sigma_0^{\rm cut}(s,q)
={2\pi\varepsilon^2\alpha^2\over s}(1-\rho)
\left[-2c_{\rm max}+4\operatorname{arctanh}c_{\rm max}
+{8\rho\over(1-\rho)^2}\operatorname{arctanh}c_{\rm max}\right] \, .
\end{equation}
Here $\hat\sigma_0^{\rm cut}(s,q)$ denotes the off-shell 2-body production cross section after the LowM angular cut, where the hat indicates that the invisible invariant mass squared $q$ is not restricted to the on-shell value $m_{A'}^2$. 
The superscript ``cut'' distinguishes it from the full-angular kernel.  
In the following we abbreviate $\sigma_0\equiv\hat\sigma_0^{\rm cut}(s,0)$ for the massless case. 
When finite masses are retained, $\sigma_0$ denotes the corresponding on-shell value $\hat\sigma_0^{\rm cut}(s,m_{A'}^2)$.  
The full-angular expression in Eq.~(\ref{eq:3_21}) instead retains the electron mass as the collinear regulator.

\begin{table}[htbp]
\centering
\begin{tabular}{c c c }
\hline
$\alpha'$ & $(\sigma_{\rm NLO}/\sigma_0-1)_{|c|<0.6}$ & $(\sigma_{\rm NLO}/\sigma_0-1)_{\rm full}$  \\
\hline
0.1 & $-0.01833$ & $-0.02867$  \\
0.3 & $-0.05503$ & $-0.08605$  \\
0.5 & $-0.09187$ & $-0.14357$  \\
\hline
\end{tabular}
\caption{
Effect of the angular cut at fixed $E_\gamma>3~\mathrm{GeV}$ for $m_{A'}=0.01~\mathrm{GeV}$, $r=1$, and $Q_S=Q_\chi=1$.
The column labeled ``full'' is integrated over the full photon angular range while retaining the same photon-energy cut. 
Its values differ from those in Fig.~\ref{fig:sec3ratio}, where $E_\gamma>1~\mathrm{GeV}$ is used.}
\label{tab:cut_effect}
\end{table}

In Tab.~\ref{tab:cut_effect} we show how the LowM cut changes the total rate.
Both columns in Tab.~\ref{tab:cut_effect} are evaluated with $E_\gamma>3~\mathrm{GeV}$, and they differ only in whether the angular requirement $|c|<0.6$ is imposed. 
The small net correction in Fig.~\ref{fig:sec3ratio} reflects a strong cancellation between the positive real-emission and negative virtual contributions.
Increasing the photon-energy threshold from $1~\GeV$ to $3~\GeV$ reduces the available real-emission phase space, while leaving the two-body Born and virtual contributions unchanged for the benchmarks considered here.
The reduced real contribution no longer fully compensates the virtual suppression, yielding $\sigma_{\rm NLO}<\sigma_0$ in Tab.~\ref{tab:cut_effect}.

The angular cut reduces the absolute Born, virtual, and real-emission rates. 
Since the virtual correction factorizes with the Born cross section, the normalized ratio $\sigma_V/\sigma_0$ is unchanged by the angular cut. 
The real-emission contribution, however, involves an off-shell invisible system with $q>0$, whose angular acceptance differs slightly from that of the on-shell Born process. 
Consequently, although the absolute real-emission rate is reduced by the angular cut, the normalized ratio $\sigma_R/\sigma_0$ is slightly enhanced. 
For all three benchmarks in Table~\ref{tab:cut_effect}, the net NLO correction remains negative with or without the angular cut.
The angular cut reduces its magnitude; for example, at $\alpha'=0.3$ the correction changes from $-8.60\%$ to $-5.50\%$, a difference of $3.10$ percentage points.
The corresponding modification of the normalized distribution shape is numerically small for the benchmark points considered here. 
In the following subsections, the LowM cuts are always imposed.

\subsection{Particle level matching \label{sec:matching}}

The matching problem considered here is specific to the following hierarchy
\begin{equation} \label{eq:6_light_hierarchy}
m_{A'}^2,\,m_s^2 \,\ll s\, .
\end{equation}
In this case we can neglect all dark-sector masses, as in the resummation of Sec.~\ref{sec:resum}.
But it should be noticed that the resummed result obtained in Sec.~\ref{sec:resum} is only valid in the domain
\begin{equation} \label{eq:6_resum_region}
m_{\rm dark}^2\ll q  \ll s\, .
\end{equation}
For $q \sim s$, we instead use the fixed-order real distribution
\begin{equation} \label{eq:6_fo_distribution}
r_{\rm FO}(q)
\equiv {1\over\sigma_0}{d\sigma_{R}^{\rm massless}\over dq}
={a_{\rm pri}\over q}
{\hat\sigma_0^{\rm cut}(s,q)\over\hat\sigma_0^{\rm cut}(s,0)}\, ,
\qquad
a_{\rm pri}={\alpha'\over\pi}
\left({Q_S^2\over12}+{Q_\chi^2\over3}\right)\, .
\end{equation}
Its singular expansion is
\begin{equation} \label{eq:6_singular_distribution}
r_{\rm sing}(q)={a_{\rm pri}\over q}\, .
\end{equation}

The corrected ``primary+secondary'' resummation result of Sec.~\ref{sec:resum_num} can be written as
\begin{equation} \label{eq:6_resummed_distribution}
r_{\rm resum}(q)
={\cal C}_{\rm sec}\,{a_{\rm pri}\over q}
\left({q\over Q_{\rm hard}^2}\right)^{a_{\rm pri}}\, ,
\end{equation}
where the secondary factor ${\cal C}_{\rm sec}$ is evaluated numerically from Eq.~(\ref{eq:5_28}).  Since ${\cal C}_{\rm sec}=1+{\cal O}(\alpha'^2)$, the first-order expansion of Eq.~(\ref{eq:6_resummed_distribution}) is precisely $r_{\rm sing}(q)$.

As we said above, $r_{\rm FO}(q)$ and $r_{\rm resum}(q)$ describe complementary kinematic regions.  
The NLO distribution retains the exact hard-region kinematics and is appropriate at large $q$, where logarithms of $q/s$ are not enhanced, but it becomes unreliable at small $q$ because the singular terms are not Sudakov suppressed.  
Conversely, the resummation derived in Sec.~\ref{sec:resum} controls the logarithmically enhanced small-$q$ region but does not retain the full nonsingular fixed-order contribution when $q$ approaches the hard scale.  
We therefore combine them with an additive matching prescription at the level of the observable~\cite{Banfi:2004yd,Bozzi:2005wk}. 
Since the two calculations share the same ${\cal O}(\alpha')$ singular term, that overlap must be subtracted to avoid double counting. 

The matched distribution is given by
\begin{equation} \label{eq:6_additive_matching}
r_{\rm match}(q)
=r_{\rm FO}(q)
+h(q)\left[r_{\rm resum}(q)-r_{\rm sing}(q)\right] \, ,
\end{equation}
with $h(q)$ a smoothing function that connects the low-$q$ and high-$q$ regions:
\begin{equation} \label{eq:6_profile}
h(q)=
\begin{cases}
1\, , & q\le q_1\, ,\\[1mm]
1-S\!\left({q-q_1\over q_2-q_1}\right)\, ,&q_1<q<q_2\, ,\\[1mm]
0\, ,&q\ge q_2\, ,
\end{cases}
\qquad S(x)=3x^2-2x^3\, ,
\end{equation}
with
\begin{equation}
q_1=0.01s=1~\GeV^2\, ,\qquad q_2=0.10s=10~\GeV^2\, .
\end{equation}

\begin{figure}[t]
\centering
\includegraphics[width=0.72\linewidth]{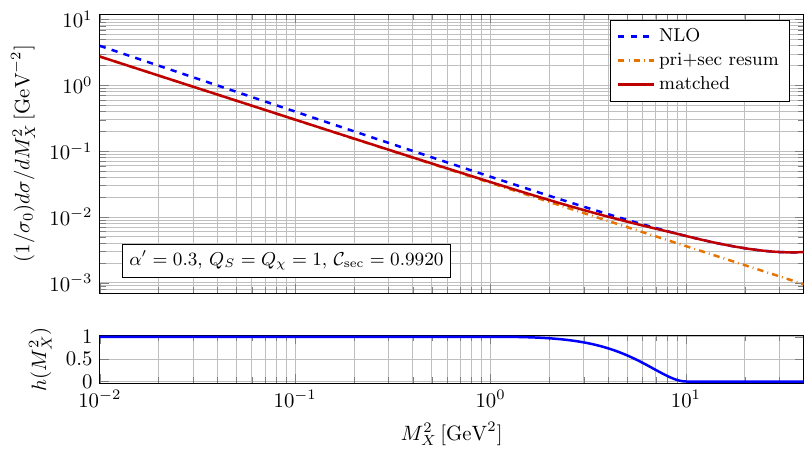}
\caption{Particle-level additive matching for $\alpha'=0.3$ and $Q_S=Q_\chi=1$.  The upper panel compares the NLO distribution, the pri+sec resummation, and the matched result.  The lower panel shows the profile function $h(q)$.  The secondary factor is ${\cal C}_{\rm sec}=0.9920$.}
\label{fig:sec6_match_diagnostic}
\end{figure}

The subtraction in Eq.~(\ref{eq:6_additive_matching}) avoids double counting and ensures
\begin{equation}
r_{\rm match}(q)=r_{\rm FO}(q)+{\cal O}(\alpha'^2)
\end{equation}
at fixed order.  In the low-$q$ region, $r_{\rm FO}-r_{\rm sing}$ is nonsingular, so the matched result approaches the resummed distribution up to a power-suppressed remainder.  In the high-$q$ region, the resummed correction is switched off and the result becomes fixed order.  
Fig.~\ref{fig:sec6_match_diagnostic} illustrates the smooth transition between these two domains.

$r_{\rm match}(q)$ in Eq.~(\ref{eq:6_additive_matching}) describes the distribution away from the resonance peak, which is not included in the continuous term.  
At a real collider one must also account for finite detector resolution.  
We introduce an auxiliary separation scale \(q_{\rm det}=0.01~\GeV^2\) and absorb the spectrum below it into an unresolved endpoint contribution. 
This scale is chosen well below the detector resolution.

Thus the complete matched particle-level distribution is
\begin{equation} \label{eq:6_complete_match}
{1\over\sigma_0}{d\sigma_{\rm match}\over dq}
=\beta_{\rm peak}\,\delta(q-m_{A'}^2)
+\Theta(q-q_{\rm det})\,r_{\rm match}(q)\, .
\end{equation}
We determine the unresolved coefficient from the inclusive NLO normalization,
\begin{equation} \label{eq:6_peak_weight}
\beta_{\rm peak}
=K_{\rm NLO}-\int_{q_{\rm det}}^{q_{\rm max}}dq\,r_{\rm match}(q)\, ,
\qquad
{\rm with}\ \ K_{\rm NLO}={\sigma_{\rm NLO}\over\sigma_0}\, .
\end{equation}
For the light-mass benchmarks, $K_{\rm NLO}$ is evaluated at $m_{A'}=0.01~\GeV$, in the stable small-mass region of the finite-mass calculation, with the same couplings, mass ratio, and cuts.
Equation~(\ref{eq:6_peak_weight}) guarantees
\begin{equation} \label{eq:6_unitarity}
\int_0^{q_{\rm max}} dq\,{d\sigma_{\rm match}\over dq}=\sigma_{\rm NLO}\, .
\end{equation}
Thus $\beta_{\rm peak}$ combines the Born-like term, virtual corrections, and radiation unresolved below $q_{\rm det}$.

\subsection{Detector level signal shape}

The distribution we obtained in the last subsection is at particle level. 
To obtain the signal shape at a real collider, we need to consider the detector effect. 
The detector level prediction is obtained by convolution with the right-tailed Crystal-Ball response used in our previous BaBar analysis~\cite{Li:2025tlg}.  
Defining
\begin{equation}
\Delta_q\equiv M_{X,{\rm obs}}^2-q\, ,\qquad
t_{\rm CB}={\Delta_q\over\sigma_{\rm CB}}\, ,
\end{equation}
we use
\begin{equation} \label{eq:6_cb_right}
R_{\rm CB}(\Delta_q)
={\cal N}_{\rm CB}
\begin{cases}
\exp(-t_{\rm CB}^2/2)\, , & t_{\rm CB}\leq\alpha_{\rm CB}\, ,\\[1.5mm]
A_{\rm CB}(B_{\rm CB}+t_{\rm CB})^{-n_{\rm CB}}\, , & t_{\rm CB}>\alpha_{\rm CB}\, ,
\end{cases}
\end{equation}
Here $\sigma_{\rm CB}$ is the Gaussian-core resolution, $\alpha_{\rm CB}>0$ is the transition point in units of $\sigma_{\rm CB}$, and $n_{\rm CB}>1$ controls the power-law falloff of the right tail.  
Since $\Delta_q=M_{X,{\rm obs}}^2-q$, positive $t_{\rm CB}$ corresponds to an upward fluctuation of the reconstructed squared missing mass, so the power-law branch at $t_{\rm CB}>\alpha_{\rm CB}$ is the right-tailed branch.  
The constants $A_{\rm CB}$ and $B_{\rm CB}$ are not independent parameters; continuity of the function and its first derivative at $t_{\rm CB}=\alpha_{\rm CB}$ fixes
\[
A_{\rm CB}=\left({n_{\rm CB}\over\alpha_{\rm CB}}\right)^{n_{\rm CB}}
\exp\!\left(-{\alpha_{\rm CB}^2\over2}\right),\qquad
B_{\rm CB}={n_{\rm CB}\over\alpha_{\rm CB}}-\alpha_{\rm CB}\, .
\]
And the normalization constant ${\cal N}_{\rm CB}$ is determined by $\int_{-\infty}^{\infty}d\Delta_q\,R_{\rm CB}(\Delta_q)=1$. 
Following Refs.~\cite{BaBar:2017tiz,Li:2025tlg}, we use $\alpha_{\rm CB}=0.765$, $n_{\rm CB}=7.54$, and $\sigma_{\rm CB}=1.5~\GeV^2$.  
Then the detector-level distribution is given by the convolution
\begin{align} \label{eq:6_detector_match}
{d\sigma_{\rm match}^{\rm det}\over dM_{X,{\rm obs}}^2} = {d\sigma_{\rm match}\over dq} \otimes R_{\rm CB}
&=\sigma_0\beta_{\rm peak}
R_{\rm CB}(M_{X,{\rm obs}}^2-m_{A'}^2)\nonumber\\
&+\sigma_0\int_{q_{\rm det}}^{q_{\rm max}}dq\,r_{\rm match}(q)
R_{\rm CB}(M_{X,{\rm obs}}^2-q)\, .
\end{align}

\begin{figure}[!htbp]
\centering
\includegraphics[width=0.8\linewidth]{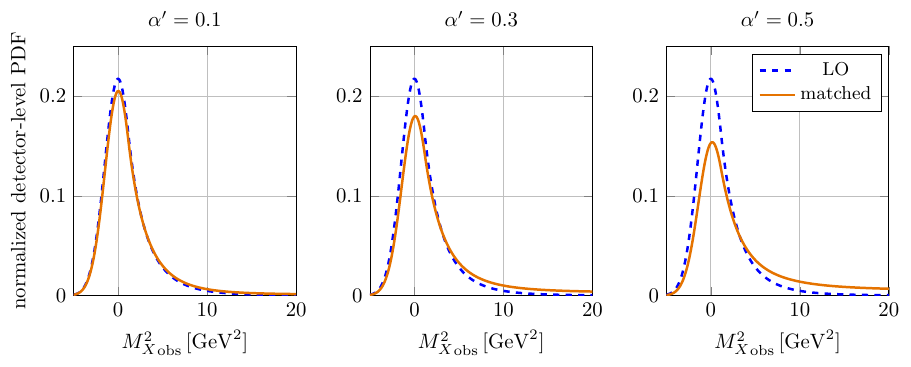}
\caption{Normalized detector-level distributions for $(Q_S,Q_\chi)=(1,1)$, $m_{A'}=0.01~\GeV$, $r=1$, and $q_{\rm det}=0.01~\GeV^2$.  The three panels correspond to $\alpha'=0.1,0.3,0.5$.  The dashed curve is the LO Crystal-Ball template and the solid curve is the matched prediction which includes the dark sector correction.}
\label{fig:sec6_match_caseA}
\end{figure}

For illustration, we take a benchmark setting as
\begin{equation} \label{eq:6_benchmarks}
m_{A'}=0.01~\GeV\, ,\qquad r=1\, ,\qquad
\alpha'=0.1,~0.3,~0.5\, ,\qquad (Q_S,Q_\chi)=(1,1)
\end{equation}
For shape comparisons, the LO and matched templates are separately normalized in the LowM fit window $-4<M_{X,{\rm obs}}^2<36~\GeV^2$.
In Fig.~\ref{fig:sec6_match_caseA} we show how the normalized detector-level distribution changes as $\alpha'$ increases. 
It is clear that the signal shape becomes broader as the dark sector correction goes stronger. 
Corresponding numerical values for different factors are listed in Table~\ref{tab:sec6_benchmark_norms}. 

\begin{table}[htbp]
\centering
\small
\begin{tabular}{ c c c c c c}
\hline
 $\alpha'$ & $Q_S$ & $Q_\chi$ & ${\cal C}_{\rm sec}$ & $\beta_{\rm peak}$ & $K_{\rm NLO}$ \\
\hline
 0.1 & 1 & 1 & $0.99884$ & $0.85956$ & $0.98167$ \\
 0.3 & 1 & 1 & $0.99201$ & $0.61354$ & $0.94497$ \\
 0.5 & 1 & 1 & $0.98181$ & $0.40534$ & $0.90813$ \\
\hline
\end{tabular}
\caption{Factors for Fig.~\ref{fig:sec6_match_caseA}.  The secondary factor ${\cal C}_{\rm sec}$ is taken from Sec.~\ref{sec:resum_num}, and $q_{\rm det}=0.01~\GeV^2$.  Here $\beta_{\rm peak}$ is the unresolved endpoint coefficient normalized to $\sigma_0$, and $K_{\rm NLO}=\sigma_{\rm NLO}/\sigma_0$.}
\label{tab:sec6_benchmark_norms}
\end{table}

\subsection{BaBar limit reinterpretation}

We finally estimate how the accepted rate and signal line shape modify the published BaBar limit.  
Let $p_i[f]$ be the probability in the $i$th observed missing-mass bin for a signal template $f$ normalized in the LowM fit window; the initial bin width is $1~\GeV^2$.
The background expectation $B_i$ is taken from the background-only fit for the $\Upsilon(3S)$ $R'_L$ LowM signal region in the supplementary material of the BaBar invisible-DP search~\cite{BaBar:2017tiz}.  
Since the public material provides the fitted curve rather than a numerical bin-by-bin workspace, we digitize the expected count in each $1~\GeV^2$ bin from that curve, following the one-region treatment in Appendix~B of Ref.~\cite{Li:2025tlg}.  
The resulting $B_i$ values sum to $129.1$ events, consistent with the 129-event sample quoted for this region.  
In the simplified recast below they are treated as fixed known background expectations.  
At a few finite-mass scan points, the truncated finite-width spectrum of Eq.~(\ref{eq:4_03}) gives negative integrals in the lowest reconstructed-mass tail bins.
For the likelihood evaluation, we merge adjacent bins until their integrated signal prediction is nonnegative, applying the same grouping to the LO template and background.
This common binning preserves the signed fit-window yield used in $K_{\rm win}$.
The affected negative weight is less than $10^{-4}$ of the accepted yield, and the resulting limit ratio differs by less than $10^{-6}$ from a comparison prescription that sets the negative bins to zero and recomputes the accepted yield.
The light-mass benchmarks in Table~\ref{tab:sec6_babar_recast} require no bin merging.
For a background-only Asimov data set and a tested signal yield $N$, the corresponding binned Poisson likelihood ratio is~\cite{Cowan:2010js}
\begin{equation} \label{eq:6_asimov_likelihood}
q_N^{\rm A}[f] = -2 \ln \frac{L(N;f)}{L(0)}
=2\sum_i\left[Np_i[f]-B_i\ln\left(1+{Np_i[f]\over B_i}\right)\right] \, .
\end{equation}
We solve Eq.~(\ref{eq:6_asimov_likelihood}) for the median expected 90\% C.L., i.e. $q_N^{\rm A}=[\Phi^{-1}(0.95)]^2\simeq2.706$ with $\Phi$ the cumulative distribution function of the standard normal distribution.
The quantity $N_{90}$ denotes the median expected upper limit on the signal yield obtained from Eq.~(\ref{eq:6_asimov_likelihood}). 
In particular, $N_{90}^{\rm LO}$ and $N_{90}^{\rm match}$ represent the expected 90\% C.L. upper limits derived from the LO and matched signal distributions, respectively. 
We then define the signal-yield ratio
\begin{equation} \label{eq:6_shape_factor}
S_{\rm fit}\equiv {N_{90}^{\rm match}\over N_{90}^{\rm LO}}\, .
\end{equation}
This quantity directly measures the impact of the shape distortion on the experimental sensitivity. 
A value of $S_{\rm fit}>1$ indicates that the matched distribution leads to a weaker constraint compared with the LO prediction, while $S_{\rm fit}<1$ corresponds to an improved sensitivity. 
For the relatively smooth background distributions considered here, the matched spectrum is generally broader than the LO one, which reduces the discrimination power of the signal shape and typically results in $N_{90}^{\rm match}>N_{90}^{\rm LO}$.

On the other hand, the accepted-rate ratio in the LowM fit-window, which is slightly different from $K_{\rm NLO}$, is given by
\begin{equation} \label{eq:6_window_rate}
K_{\rm win}={\displaystyle\int_{-4~\GeV^2}^{36~\GeV^2}dM_{X,{\rm obs}}^2\,
{d\sigma_{\rm match}^{\rm det}\over dM_{X,{\rm obs}}^2}
\over\displaystyle\int_{-4~\GeV^2}^{36~\GeV^2}dM_{X,{\rm obs}}^2\,
{d\sigma_{\rm LO}^{\rm det}\over dM_{X,{\rm obs}}^2}}\, .
\end{equation}
Since the production rate is proportional to $\varepsilon^2$, we use the ratio of expected sensitivities to rescale the published limit. The resulting benchmark estimate, denoted by $\varepsilon'$, is 
\begin{equation} \label{eq:6_modified_limit}
\varepsilon' = \varepsilon_0 \sqrt{ {S_{\rm fit}\over K_{\rm win}} }\, .
\end{equation}
Here $\varepsilon_0$ is the published BaBar limit~\cite{BaBar:2017tiz}. 
Equation~(\ref{eq:6_modified_limit}) transfers a relative expected-sensitivity change from the simplified likelihood to that reference curve. 

\begin{table}[htbp]
\centering
\small
\begin{tabular}{c c c c  c}
\hline
$(Q_S,Q_\chi)$ & $\alpha'$ & $S_{\rm fit}$ & $K_{\rm win}$ & $\varepsilon'/\varepsilon_0$ \\
\hline
(1,1) & 0.1 & $1.0453$ & $0.9770$ &  $1.0343$ \\
(1,1) & 0.3 & $1.1431$ & $0.9310$ &  $1.1081$ \\
(1,1) & 0.5 & $1.2462$ & $0.8848$ &  $1.1868$ \\
(1,1/4) & 0.1 & $1.0111$ & $0.9923$ &  $1.0094$ \\
(1,1/4) & 0.3 & $1.0339$ & $0.9768$ &  $1.0288$ \\
(1,1/4) & 0.5 & $1.0576$ & $0.9613$ &  $1.0489$ \\
(1/4,1) & 0.1 & $1.0365$ & $0.9833$ &  $1.0267$ \\
(1/4,1) & 0.3 & $1.1140$ & $0.9499$ &  $1.0829$ \\
(1/4,1) & 0.5 & $1.1956$ & $0.9163$ &  $1.1423$ \\
\hline
\end{tabular}
\caption{Shape, accepted-rate, and limit factors at $m_{A'}=0.01~\GeV$.  The shape factor is obtained from the binned Poisson Asimov likelihood using the digitized $\Upsilon(3S)$ $R'_L$ background expectation described in the text. The definitions of $K_{\rm win}$ and $\varepsilon'/\varepsilon_0$ can be found in the text.}
\label{tab:sec6_babar_recast}
\end{table}

\begin{figure}[!htbp]
\centering

\subfloat[
    $(Q_S,Q_\chi)=(1,1)$.
    \label{fig:sec6_babar_limit_A}
]{%
    \includegraphics[width=0.6\linewidth]
    {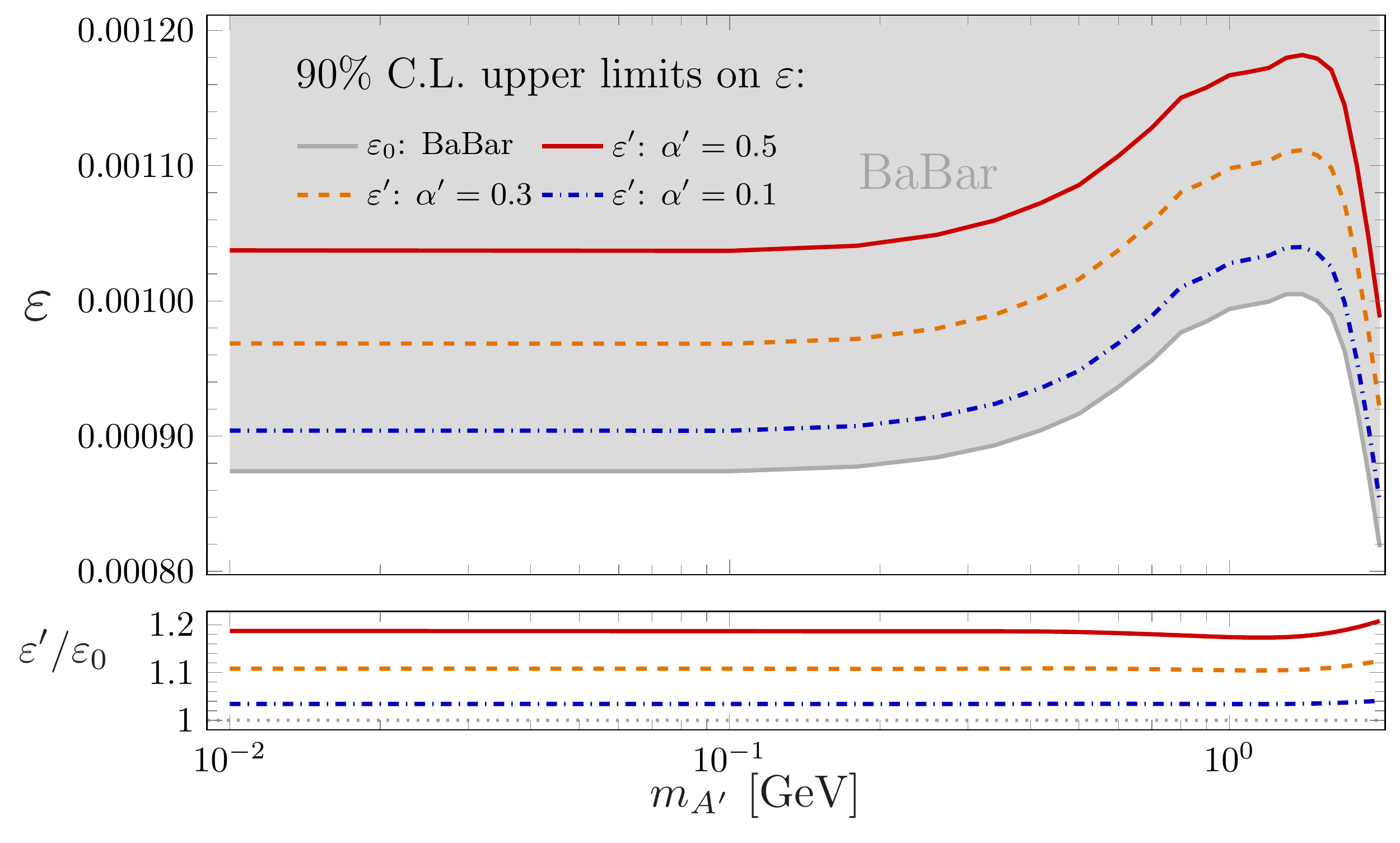}%
}

\par\vspace{0.2em}

\subfloat[
    $(Q_S,Q_\chi)=(1,1/4)$.
    \label{fig:sec6_babar_limit_B}
]{%
    \includegraphics[width=0.60\linewidth]
    {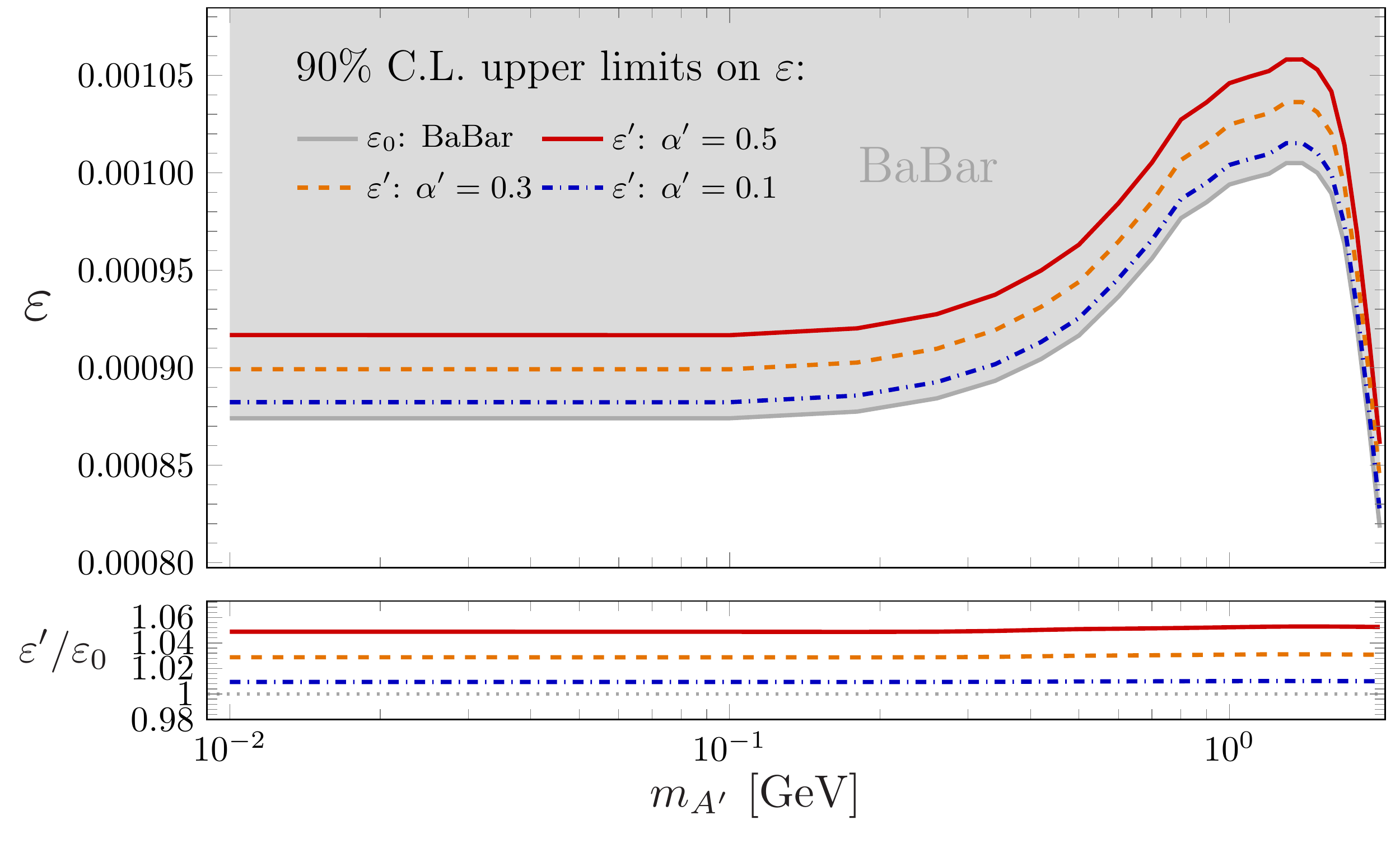}%
}

\par\vspace{0.2em}

\subfloat[
    $(Q_S,Q_\chi)=(1/4,1)$.
    \label{fig:sec6_babar_limit_C}
]{%
    \includegraphics[width=0.6\linewidth]
    {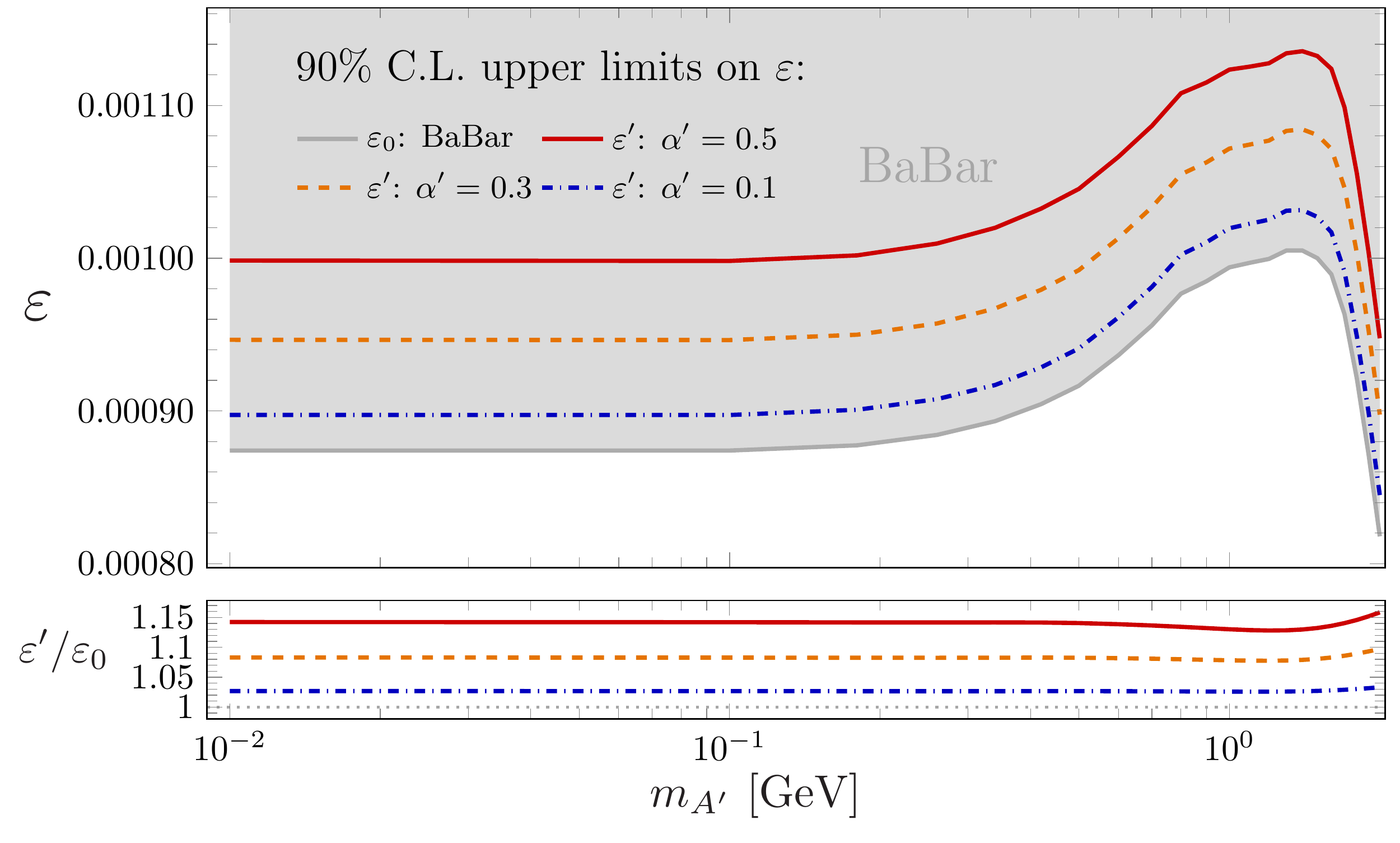}%
}

\caption{
The published BaBar 90\% C.L. upper limit on $\varepsilon$ and the
modified limits obtained in our simplified recast.
Panels (a), (b), and (c) correspond to
$(Q_S,Q_\chi)=(1,1)$, $(1,1/4)$, and $(1/4,1)$, respectively.
In each panel, the lower inset shows the ratio of the modified limit
to the published BaBar limit.
}
\label{fig:sec6_babar_limits}
\end{figure}

Table~\ref{tab:sec6_babar_recast} gives the factors at $m_{A'}=0.01~\GeV$, evaluated with the light-mass matched spectrum. 
For $Q_S=Q_\chi=1$ and $\alpha'=0.3$, the accepted-rate factor is $K_{\rm win}=0.9310$ and the shape factor is $S_{\rm fit}=1.1431$, giving $\varepsilon'/\varepsilon_0=1.1081$.
The accepted yield is reduced by $6.90\%$; a rate-only reinterpretation would weaken the limit by $3.64\%$, while including the shape distortion increases the weakening to $10.81\%$.
Both accepted-rate suppression and shape broadening therefore weaken the limit in this benchmark.
For $(Q_S,Q_\chi)=(1,1)$, the limits are weakened by $3.43\%$, $10.81\%$, and $18.68\%$ for $\alpha'=0.1$, $0.3$, and $0.5$, respectively.
For $(Q_S,Q_\chi)=(1,1/4)$, the corresponding weakenings are $0.94\%$, $2.88\%$, and $4.89\%$.
For $(Q_S,Q_\chi)=(1/4,1)$, they are $2.67\%$, $8.29\%$, and $14.23\%$.
All nine light-mass benchmarks in Table~\ref{tab:sec6_babar_recast} have $K_{\rm win}<1$ and $S_{\rm fit}>1$.

To obtain the modified BaBar limit for both massless and finite-mass cases, we use the massless matched spectrum of Eq.~(\ref{eq:6_complete_match}) for $m_{A'}\leq0.1~\GeV$ and the finite-mass NLO distribution of Sec.~\ref{sec:wo_res} for $m_{A'}\geq0.5~\GeV$.  
In the intermediate region $0.1<m_{A'}<0.5~\GeV$, the result is obtained by a smooth interpolation between these two limiting cases.
Figure~\ref{fig:sec6_babar_limits} displays the resulting sensitivity-rescaled curves for the different charge assignments. 
The largest shift among the displayed scan points is a $20.77\%$ weakening, attained at $m_{A'}=2~\GeV$ for $(Q_S,Q_\chi)=(1,1)$ and $\alpha'=0.5$.
The size of the shift depends on the mass, coupling, and charges.

Future lepton colliders with sufficiently good photon-energy resolution could probe dark-sector dynamics through precise measurements of the $M_X^2$ spectrum. 
Resolving the radiative tail would provide sensitivity to the dark charges and couplings, and thus reveal the dynamics in the dark sector. 
In the BaBar analysis considered here, however, the finite photon-energy resolution smears these spectral features. 
After including this effect, our simplified recast finds changes in the exclusion limit on $\varepsilon$ reaching the $\mathcal{O}(10\%)$ level in the benchmarks studied.

\section{Conclusions \label{sec:conclu}}

We have developed an analytic framework for invisible dark sector radiation in monophoton searches at electron--positron colliders.
In a dark Abelian Higgs model with a light dark fermion, we connect the inclusive NLO rate, the resummed endpoint spectrum, and detector-level signal shapes. 
This connection makes the dependence on dark sector charges and couplings explicit and provides an analytic benchmark for Monte Carlo dark shower study.

The neutral initiating dark photon determines the primary logarithmic structure: its splitting kernels have no soft endpoint poles and generate single collinear logarithms. 
Within our fixed-coupling, leading-power, angular-ordered treatment, coherent secondary radiation contributes an $M_X^2$-independent factor $\mathcal{C}_{\rm sec}$ and leaves the endpoint power unchanged.
This result identifies the radiation mechanism controlling the missing-mass spectrum when the entire dark final state is invisible.

Applying the matched spectra to a simplified BaBar reinterpretation shows that kinetic-mixing constraints depend on both the accepted signal rate and the spectral shape. 
Their combined treatment is therefore essential for translating monophoton searches into constraints on an interacting dark sector. 
More broadly, the framework connects dark sector dynamics to a measurable recoil observable, and provides a basis for assessing how improved missing-mass resolution could help to probe interactions in the invisible world.

\acknowledgments
We appreciate helpful discussions with Hai-Tao Li and Shao-Feng Ge. 
This work was supported by the National Natural Science Foundation of China (NSFC) under grant Nos. 12105118 and 11947118.
OpenAI ChatGPT was used to assist with coding and numerical calculations under the authors' supervision.

\bibliographystyle{JHEP}
\bibliography{ref}

\end{document}